\documentclass{llncs}
\usepackage{multicol}

\usepackage{amsmath}
\usepackage{amsfonts}
\usepackage{amssymb}
\usepackage{verbatim}

\usepackage{graphicx}
\usepackage{epic}
\usepackage{eepic}
\usepackage{epsfig,float}

\usepackage{multicol}
\renewcommand{\le}{\leqslant}
\renewcommand{\ge}{\geqslant}

\newcommand{\eps}{\varepsilon}
\newcommand{\emp}{\emptyset}

\newcommand{\Sig}{\Sigma}

\newcommand{\noin}{\noindent}

\newcommand{\bi}{\begin{itemize}}
\newcommand{\ei}{\end{itemize}}
\newcommand{\be}{\begin{enumerate}}
\newcommand{\ee}{\end{enumerate}}
\newcommand{\bd}{\begin{description}}
\newcommand{\ed}{\end{description}}
\newcommand{\bq}{\begin{quote}}
\newcommand{\eq}{\end{quote}}

\newcommand{\der}[1]{\xrightarrow{#1}}

\newcommand{\cD}{{\mathcal D}}

\newcommand{\cK}{{\mathcal K}}
\newcommand{\cL}{{\mathcal L}}

\newcommand{\cN}{{\mathcal N}}
\newcommand{\cP}{{\mathcal P}}

\newcommand{\cR}{{\mathcal R}}

\newcommand{\cU}{{\mathcal U}}
\newcommand{\cV}{{\mathcal V}}
\newcommand{\one}{{\mathbf 1}}

\newcommand{\bs}{\setminus}

\newcommand{\rev}{\mathbb{R}}

\title{Universal Witnesses for State Complexity of Basic Operations Combined with Reversal\thanks{This work was supported by the Natural Sciences and Engineering Research Council of Canada under grant No.~OGP0000871.}
}
\author{Janusz~Brzozowski and David Liu}

\authorrunning{Brzozowski, Liu}   

\institute{David R. Cheriton School of Computer Science, University of Waterloo, \\
Waterloo, ON, Canada N2L 3G1\\
{\tt \{brzozo,dyliu\}@uwaterloo.ca}
}
\begin{document}
\maketitle
\begin{abstract}
We study the state complexity of boolean operations, concatenation and star with 
one or two of the argument languages reversed.
We derive tight upper bounds for the symmetric differences and differences of such languages. 
We prove that the previously discovered bounds for union, intersection, concatenation and star of such languages can all be met by the recently introduced
universal witnesses and their variants. 
\smallskip

\noin
\end{abstract}

\section{Introduction}
For background on state complexity see~\cite{Brz10,Brz12,Yu01}.
The \emph{state complexity of a regular language} is the number of states in the minimal deterministic finite automaton (DFA) recognizing the language.
The \emph{state complexity of an operation} on regular languages is the worst-case state complexity of the result of the operation as a function of the state complexities of the arguments.

The state complexity of basic operations combined with reversal was studied in 2008 by Liu, Martin-Vide, A.~Salomaa, and Yu~\cite{LMSY_IC08}. Let $K$ and $L$ be two regular languages over alphabet $\Sig$, and let their state complexities be $m$ and $n$, respectively.
The basic operations considered in~\cite{LMSY_IC08} were union ($K\cup L$), intersection ($K\cap L$), product (catenation or concatenation) ($KL$) and star ($L^*$), and
reversal ($L^R$) was added to these operations.
It was shown that $(2^m-1)(2^n-1)+1$ is a tight upper bound for  $(K\cup L)^R=K^R\cup L^R$ and
$(K\cap L)^R=K^R\cap L^R$. 
It was also proved that $3\cdot 2^{m+n-2} - (2^n-1)$ is an upper bound for 
$(KL)^R=L^RK^R$, but the question of tightness was left open.
Cui, Gao, Kari and Yu~\cite{CGKY_TCS12} answered this question positively, and also showed that $3\cdot 2^{m+n-2}$ is an upper bound for $K^RL$.
In another paper~\cite{CGKY_IJFCS12}, they proved that $(m-1)2^n+2^{n-1}-(m-1)$ is a tight upper bound for $KL^R$.
Gao, K. Salomaa, and Yu~\cite{GSY_FI08} demonstrated that $2^n$ is a tight upper bound for $(L^*)^R=(L^R)^*$. 
Gao and Yu~\cite{GY_FI12}  found the tight upper bound $m2^n-(m-1)$ for $K\cup L^R$ and $K\cap L^R$.
Thus  eight basic operations with reversal added have been considered so far.

There are two steps in finding the state complexity of an operation:
one has to establish an upper bound for this complexity, and then find  languages to act as witnesses to show that the  bound is tight. 
One usually defines a sequence $(L_n\mid n\ge k)$ of languages, where  $k$ is some small positive integer.  This sequence will be called a \emph{stream} of languages; for example, $(\{a,b\}^*a\{a,b\}^{n-3} \mid n\ge 3)$ is a stream.
The languages  in a stream normally differ only in the parameter $n$. 
Usually, two different streams have been used as witnesses for binary operations.

In 2012, Brzozowski~\cite{Brz12} defined the notion of \emph{permutational equivalence}.
Two languages $K$ and $L$ over $\Sig$ are permutationally equivalent if one can be obtained from the other by permuting the letters of the alphabet. 
For example, $K=\{a,b\}^*a\{a,b\}^{n-3}$ is permutationally equivalent to 
$L=\{a,b\}^*b\{a,b\}^{n-3}$.
These two languages have the same properties,  only the letters have been renamed.

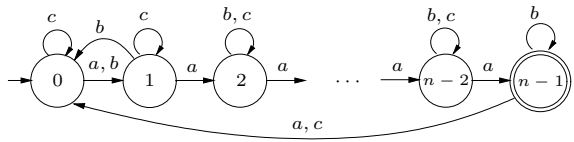
\begin{figure}[hbt]
\begin{center}
\setlength{\unitlength}{0.00043745in}
\begingroup\makeatletter\ifx\SetFigFont\undefined%
\gdef\SetFigFont#1#2#3#4#5{%
  \reset@font\fontsize{#1}{#2pt}%
  \fontfamily{#3}\fontseries{#4}\fontshape{#5}%
  \selectfont}%
\fi\endgroup%
{\renewcommand{\dashlinestretch}{30}
\begin{picture}(6677,1617)(0,-10)
\put(469,1401){\makebox(0,0)[lb]{\smash{{\SetFigFont{7}{8.4}{\familydefault}{\mddefault}{\updefault}$c$}}}}
\put(2747.000,1151.333){\arc{333.333}{2.2143}{7.2105}}
\blacken\path(2884.638,1114.417)(2847.000,1018.000)(2925.107,1085.913)(2884.638,1114.417)
\put(1667.000,1151.333){\arc{333.333}{2.2143}{7.2105}}
\blacken\path(1804.638,1114.417)(1767.000,1018.000)(1845.107,1085.913)(1804.638,1114.417)
\put(5179.000,1158.333){\arc{333.333}{2.2143}{7.2105}}
\blacken\path(5316.638,1121.417)(5279.000,1025.000)(5357.107,1092.913)(5316.638,1121.417)
\put(6312.000,1211.333){\arc{333.333}{2.2143}{7.2105}}
\blacken\path(6449.638,1174.417)(6412.000,1078.000)(6490.107,1145.913)(6449.638,1174.417)
\put(591,703){\ellipse{630}{630}}
\put(2755,703){\ellipse{630}{630}}
\put(1685,701){\ellipse{630}{630}}
\put(5187,703){\ellipse{630}{630}}
\put(6309,696){\ellipse{720}{720}}
\put(6309,697){\ellipse{630}{630}}
\path(12,703)(282,703)
\blacken\path(162.000,673.000)(282.000,703.000)(162.000,733.000)(162.000,673.000)
\path(1992,703)(2442,703)
\blacken\path(2322.000,673.000)(2442.000,703.000)(2322.000,733.000)(2322.000,673.000)
\path(3072,703)(3522,703)
\blacken\path(3402.000,673.000)(3522.000,703.000)(3402.000,733.000)(3402.000,673.000)
\path(912,703)(1362,703)
\blacken\path(1242.000,673.000)(1362.000,703.000)(1242.000,733.000)(1242.000,673.000)
\path(4422,711)(4872,711)
\blacken\path(4752.000,681.000)(4872.000,711.000)(4752.000,741.000)(4752.000,681.000)
\path(5495,703)(5945,703)
\blacken\path(5825.000,673.000)(5945.000,703.000)(5825.000,733.000)(5825.000,673.000)
\path(1497,965)(1496,967)(1493,970)
	(1488,977)(1480,986)(1469,998)
	(1456,1014)(1441,1031)(1424,1049)
	(1405,1068)(1385,1088)(1364,1106)
	(1341,1125)(1318,1141)(1292,1157)
	(1265,1170)(1236,1182)(1205,1191)
	(1172,1196)(1137,1198)(1103,1195)
	(1070,1188)(1039,1178)(1011,1165)
	(986,1150)(962,1134)(940,1116)
	(919,1097)(900,1077)(881,1057)
	(864,1037)(849,1018)(835,1000)
	(824,985)(814,972)(799,950)
\blacken\path(841.814,1066.047)(799.000,950.000)(891.387,1032.247)(841.814,1066.047)
\path(5997,493)(5996,493)(5995,492)
	(5993,491)(5989,489)(5983,487)
	(5976,483)(5966,479)(5954,474)
	(5940,468)(5923,460)(5903,452)
	(5881,442)(5856,432)(5829,420)
	(5799,408)(5766,395)(5731,381)
	(5695,367)(5656,352)(5615,337)
	(5572,321)(5527,305)(5481,289)
	(5433,273)(5383,257)(5332,241)
	(5280,226)(5225,210)(5169,195)
	(5111,180)(5051,165)(4989,151)
	(4925,137)(4859,123)(4789,110)
	(4717,98)(4642,86)(4564,75)
	(4482,64)(4397,54)(4308,45)
	(4215,37)(4118,30)(4019,24)
	(3915,19)(3810,15)(3702,13)
	(3605,12)(3509,13)(3412,15)
	(3317,17)(3223,21)(3131,25)
	(3041,30)(2952,37)(2866,43)
	(2782,51)(2699,59)(2619,67)
	(2540,76)(2463,86)(2387,96)
	(2313,106)(2240,117)(2169,128)
	(2099,139)(2030,151)(1962,163)
	(1895,175)(1829,188)(1764,200)
	(1701,213)(1638,226)(1577,239)
	(1517,252)(1459,265)(1402,277)
	(1347,290)(1294,302)(1242,314)
	(1194,326)(1147,337)(1103,348)
	(1062,358)(1023,367)(988,376)
	(956,384)(926,391)(900,398)
	(877,404)(858,409)(841,413)
	(827,417)(816,420)(807,422)
	(801,424)(792,426)
\blacken\path(915.650,429.254)(792.000,426.000)(902.635,370.683)(915.650,429.254)
\put(536,636){\makebox(0,0)[lb]{\smash{{\SetFigFont{7}{8.4}{\rmdefault}{\mddefault}{\updefault}$0$}}}}
\put(1624,636){\makebox(0,0)[lb]{\smash{{\SetFigFont{7}{8.4}{\rmdefault}{\mddefault}{\updefault}$1$}}}}
\put(2711,636){\makebox(0,0)[lb]{\smash{{\SetFigFont{7}{8.4}{\rmdefault}{\mddefault}{\updefault}$2$}}}}
\put(2120,801){\makebox(0,0)[lb]{\smash{{\SetFigFont{7}{8.4}{\familydefault}{\mddefault}{\updefault}$a$}}}}
\put(3192,823){\makebox(0,0)[lb]{\smash{{\SetFigFont{7}{8.4}{\familydefault}{\mddefault}{\updefault}$a$}}}}
\put(5622,823){\makebox(0,0)[lb]{\smash{{\SetFigFont{7}{8.4}{\familydefault}{\mddefault}{\updefault}$a$}}}}
\put(957,838){\makebox(0,0)[lb]{\smash{{\SetFigFont{7}{8.4}{\familydefault}{\mddefault}{\updefault}$a,b$}}}}
\put(4917,658){\makebox(0,0)[lb]{\smash{{\SetFigFont{6}{7.2}{\familydefault}{\mddefault}{\updefault}$n-2$}}}}
\put(3875,644){\makebox(0,0)[lb]{\smash{{\SetFigFont{7}{8.4}{\familydefault}{\mddefault}{\updefault}$\cdots$}}}}
\put(4550,816){\makebox(0,0)[lb]{\smash{{\SetFigFont{7}{8.4}{\familydefault}{\mddefault}{\updefault}$a$}}}}
\put(6184,1431){\makebox(0,0)[lb]{\smash{{\SetFigFont{7}{8.4}{\familydefault}{\mddefault}{\updefault}$b$}}}}
\put(3364,133){\makebox(0,0)[lb]{\smash{{\SetFigFont{7}{8.4}{\familydefault}{\mddefault}{\updefault}$a,c$}}}}
\put(6040,640){\makebox(0,0)[lb]{\smash{{\SetFigFont{6}{7.2}{\familydefault}{\mddefault}{\updefault}$n-1$}}}}
\put(4962,1408){\makebox(0,0)[lb]{\smash{{\SetFigFont{7}{8.4}{\familydefault}{\mddefault}{\updefault}$b,c$}}}}
\put(1039,1273){\makebox(0,0)[lb]{\smash{{\SetFigFont{7}{8.4}{\familydefault}{\mddefault}{\updefault}$b$}}}}
\put(2547,1423){\makebox(0,0)[lb]{\smash{{\SetFigFont{7}{8.4}{\familydefault}{\mddefault}{\updefault}$b,c$}}}}
\put(1557,1423){\makebox(0,0)[lb]{\smash{{\SetFigFont{7}{8.4}{\familydefault}{\mddefault}{\updefault}$c$}}}}
\put(592.000,1136.333){\arc{333.333}{2.2143}{7.2105}}
\blacken\path(729.638,1099.417)(692.000,1003.000)(770.107,1070.913)(729.638,1099.417)
\end{picture}
}
\end{center}
\caption{DFA $\cU_n(a,b,c)$ of a complex language $U_n(a,b,c)$.} 
\label{fig:witness}
\end{figure}

The DFA 
$\cU_n(a,b,c)=(Q,\Sig,\delta,q_0, F)$ of Fig.~\ref{fig:witness} and its language, $U_n(a,b,c)$,  were proposed in~\cite{Brz12} as the ``universal witness'' DFA and language, for $n\ge 3$. 
The permutationally equivalent language and DFA of $U_n(a,b,c)$ and 
$\cU_n(a,b,c)$ obtained by interchanging $a$ and $b$
are denoted by $U_n(b,a,c)$ and  $\cU_n(b,a,c)$.
The restriction of the language and  the DFA to alphabet $\{a,b\}$ is denoted by
$U_n(a,b,\emp)$ and  $\cU_n(a,b,\emp)$.

It was proved in~\cite{Brz12} that the bound $2^{n-1}+2^{n-2}$ for star is met by $U_n(a,b,\emp)$, and the bound $(m-1)2^n+2^{n-1}$ for product, by $U_m(a,b,c)U_n(a,b,c)$.
The bound $mn$ for union, intersection, difference ($K\setminus L$) and symmetric difference
($K\oplus L$) is met by two permutationally equivalent streams 
$(U_m(a,b,c) \mid m \ge 3)$ and $(U_n(b,a,c) \mid n \ge 3)$.
Thus $U_n(a,b,c)$ is a universal witness for the basic operations.

The inputs to the DFA $\cU_n(a,b,c)$ perform the following transformations on the set 
$Q=\{0,\ldots,n-1\}$ of states.
Input $a$ is a \emph{cycle} of all $n$ states, and this is denoted by $a:(0,\ldots,n-1)$.
Input $b$ is a \emph{transposition} of  0 and 1, and does not affect any other states; this is denoted by
$b:(0,1)$. 
Input $c$ is a \emph{singular} transformation sending state $n-1$ to state 0, and not affecting any other states; it is denoted by $c:{n-1\choose 0}$.
It is known~\cite{Brz12}  that the inputs of $\cU_n(a,b,c)$ of Fig.~\ref{fig:witness} perform all $n^n$ transformations of states, and also that the state complexity of the reverse of $U_n(a,b,c)$
is $2^n$; the latter result follows by a theorem from~\cite{SWY04}.

A \emph{dialect} of $U_n(a,b,c)$ is the language of  any DFA with three inputs $a$, $b$, and $c$, where $a$ is a cycle as above, $b$ is the transposition of \emph{any} two states $(p,q)$, and 
$c$ is a singular transformation ${r \choose s}$ sending \emph{any} state $r$ to \emph{any} state $s\neq r$.
The initial state is always 0, but the set of final states is arbitrary, as long as the DFA is minimal.

The universal witness and the notion of dialect have been extended to quaternary alphabets~\cite{Brz12}, by adding a fourth input $d$ which performs the identity permutation, denoted by $d:\one_Q$.
The concepts of permutational equivalence and dialect are extended in the obvious way to quaternary languages and DFA's.

In this paper, we extend the notion of basic operations from~\cite{LMSY_IC08} by including difference and symmetric difference. 
Altogether, we study the following 13 languages with these basic operations and reversal:
\goodbreak

\noin
$\hspace{2cm} K\cup L^R, \quad K\cap L^R, \quad K\setminus L^R,  \quad K \oplus L^R, 
\quad L^R\setminus K,$ \\ \mbox{}
$\hspace{2cm} K^R \cup L^R, \quad K^R \cap L^R, \quad K^R \setminus L^R, \quad K^R \oplus L^R,$\\  \mbox{}
$\hspace{2cm} KL^R, \quad K^RL, \quad K^RL^R$ and  $(K^R)^*$.
\smallskip

Our contributions are as follows:
\be
\item
We prove the conjecture from~\cite{Brz12} that the bound $mn$ for all four boolean operations in the case where $m\neq n$ is met by two identical streams of languages $U_m(a,b,\emp)$ and
$U_n(a,b,\emp)$.
\item
We derive the bound $m2^n-(m-1)$ for $K_m\setminus L_n^R$ and $L_n^R\setminus K_m$ and the bound $m2^n$ for $K_m\oplus L_n^R$, and  show that these bounds and the known bounds for 
$K_m\cup L_n^R$ and $K_m\cap L_n^R$ are met by two identical streams of languages $U_m(a,b,c)$ and $U_n(a,b,c)$.
This reduces the size of the alphabet for union and intersection from four in~\cite{GY_FI12} to three.
\item
We derive the bound $(2^m-1)(2^n-1)+1$ for $K_m^R\setminus L_n^R$, and the bound $2^{m+n-1}$ for $K_m^R\oplus L_n^R$, and  show that these bounds and the known bounds for 
$K_m^R\cup L_n^R$ and $K_m^R\cap L_n^R$ are met by two streams,  $U_{\{0,2\},m}(a,b,c)$ and $U_{\{1,3\},n}(b,a,c)$, where the set of final states
in $\cU_{\{0,2\},m}(a,b,c)$ (respectively, $\cU_{\{1,3\},n}(b,a,c)$) is $\{0,2\}$ (respectively $\{1,3\}$).
\item
We prove that the known bound for 
$K_m L_n^R$ is met by two identical streams of languages $U_m(a,b,c)$ and $U_n(a,b,c)$.
\item
We show that the known bound for 
$K_m^R L_n$ is met by two permutationally equivalent  dialects of $U_n(a,b,c,d)$.
\item
We prove that the known bound for 
$(K_mL_n)^R=L_n^RK_m^R$ is met by two permutationally equivalent streams $(U_m(a,b,c,d)\mid m\ge 3)$ and $(U_n(d,c,b,a)\mid n\ge 3)$.
Our proof is considerably simpler than the one in~\cite{CGKY_TCS12}.
\item
We note that the original proof in~\cite{GSY_FI08} uses a dialect of $U_n(a,b,c)$, and point out that the known bound is met by $U_n(a,b,c)$ with final state 0.
\item 
In obtaining the results above, we prove Conjectures 1--4, 8, 11, and 14 of~\cite{Brz12}.
\ee

The remainder of the paper is structured as follows.
In Section~\ref{sec:norev} we deal with boolean operations with no reversed arguments.
Boolean operations with one and two reversed arguments are considered in Sections~\ref{sec:onerev} and~\ref{sec:tworev}.
Product and star and examined in Section~\ref{sec:prodstar}, and Section~\ref{sec:conc}
concludes the paper.

\section{Boolean Operations with No Reversed Arguments}
 \label{sec:norev}
 Let $K\circ L$ denote any one of the four boolean operations $K\cup L$, $K\cap L$, 
 $K\oplus L$ and $K\setminus L$.
 It is well-known that, if $m$ and $n$ are the state complexities of $K$ and $L$, the state complexity of 
 $K\circ L$ is less than or equal to $mn$.
It was shown in~\cite{Brz12} that $U_m(a,b,\emp)$ and $U_n(b,a,\emp)$ are witnesses to this bound, and it was conjectured that  
 $U_m(a,b,\emp)$ and $U_n(a,b,\emp)$ are also witnesses if $m\neq n$.  
 We now prove this conjecture.
 The DFA's $\cD_1=\cU_4(a,b,\emp)$ and $\cD_2=\cU_6(a,b,\emp)$ are shown in Fig.~\ref{fig:binarykk}. Their direct product, $\cP$,
 shown in Fig.~\ref{fig:boolkk},  serves as a basis for all four cases. 

\begin{figure}[bht]
\begin{center}
\setlength{\unitlength}{0.00039370in}
\begingroup\makeatletter\ifx\SetFigFont\undefined%
\gdef\SetFigFont#1#2#3#4#5{%
  \reset@font\fontsize{#1}{#2pt}%
  \fontfamily{#3}\fontseries{#4}\fontshape{#5}%
  \selectfont}%
\fi\endgroup%
{\renewcommand{\dashlinestretch}{30}
\begin{picture}(12241,2324)(0,-10)
\put(8495,109){\makebox(0,0)[lb]{\smash{{\SetFigFont{9}{10.8}{\rmdefault}{\mddefault}{\updefault}$\cD_2=\cU_6(a,b,\emp)$}}}}
\put(4211.500,1735.929){\arc{394.717}{2.4948}{6.9299}}
\blacken\thicklines
\path(4371.033,1757.097)(4369.000,1617.000)(4442.998,1735.977)(4371.033,1757.097)
\thinlines
\put(8554.500,1735.929){\arc{394.717}{2.4948}{6.9299}}
\blacken\thicklines
\path(8714.033,1757.097)(8712.000,1617.000)(8785.998,1735.977)(8714.033,1757.097)
\thinlines
\put(9672.500,1750.929){\arc{394.717}{2.4948}{6.9299}}
\blacken\thicklines
\path(9832.033,1772.097)(9830.000,1632.000)(9903.998,1750.977)(9832.033,1772.097)
\thinlines
\put(10797.500,1735.929){\arc{394.717}{2.4948}{6.9299}}
\blacken\thicklines
\path(10957.033,1757.097)(10955.000,1617.000)(11028.998,1735.977)(10957.033,1757.097)
\thinlines
\put(11914.500,1757.929){\arc{394.717}{2.4948}{6.9299}}
\blacken\thicklines
\path(12074.033,1779.097)(12072.000,1639.000)(12145.998,1757.977)(12074.033,1779.097)
\thinlines
\put(4194,1319){\ellipse{630}{630}}
\put(3065,1331){\ellipse{630}{630}}
\put(8552,1325){\ellipse{630}{630}}
\put(9673,1332){\ellipse{630}{630}}
\put(6402,1317){\ellipse{630}{630}}
\put(7459,1335){\ellipse{630}{630}}
\put(10787,1324){\ellipse{630}{630}}
\put(847,1300){\ellipse{630}{630}}
\put(1930,1295){\ellipse{630}{630}}
\put(4195,1316){\ellipse{540}{540}}
\put(11918,1330){\ellipse{630}{630}}
\put(11911,1335){\ellipse{540}{540}}
\path(2172,1519)(2802,1519)
\blacken\thicklines
\path(2667.000,1481.500)(2802.000,1519.000)(2667.000,1556.500)(2667.000,1481.500)
\thinlines
\path(3328,1534)(3958,1534)
\blacken\thicklines
\path(3823.000,1496.500)(3958.000,1534.000)(3823.000,1571.500)(3823.000,1496.500)
\thinlines
\path(1047,1526)(1677,1526)
\blacken\thicklines
\path(1542.000,1488.500)(1677.000,1526.000)(1542.000,1563.500)(1542.000,1488.500)
\thinlines
\path(1685,1114)(1100,1114)
\blacken\thicklines
\path(1235.000,1151.500)(1100.000,1114.000)(1235.000,1076.500)(1235.000,1151.500)
\thinlines
\path(7243,1115)(6658,1115)
\blacken\thicklines
\path(6793.000,1152.500)(6658.000,1115.000)(6793.000,1077.500)(6793.000,1152.500)
\thinlines
\path(7721,1550)(8306,1550)
\blacken\thicklines
\path(8171.000,1512.500)(8306.000,1550.000)(8171.000,1587.500)(8171.000,1512.500)
\thinlines
\path(8817,1542)(9402,1542)
\blacken\thicklines
\path(9267.000,1504.500)(9402.000,1542.000)(9267.000,1579.500)(9267.000,1504.500)
\thinlines
\path(9942,1542)(10527,1542)
\blacken\thicklines
\path(10392.000,1504.500)(10527.000,1542.000)(10392.000,1579.500)(10392.000,1504.500)
\thinlines
\path(6627,1557)(7212,1557)
\blacken\thicklines
\path(7077.000,1519.500)(7212.000,1557.000)(7077.000,1594.500)(7077.000,1519.500)
\thinlines
\path(5577,1324)(6064,1324)
\blacken\thicklines
\path(5944.000,1294.000)(6064.000,1324.000)(5944.000,1354.000)(5944.000,1294.000)
\thinlines
\path(12,1317)(499,1317)
\blacken\thicklines
\path(379.000,1287.000)(499.000,1317.000)(379.000,1347.000)(379.000,1287.000)
\thinlines
\path(11044,1549)(11629,1549)
\blacken\thicklines
\path(11494.000,1511.500)(11629.000,1549.000)(11494.000,1586.500)(11494.000,1511.500)
\thinlines
\path(3949,1084)(3948,1084)(3946,1083)
	(3943,1081)(3937,1079)(3929,1075)
	(3918,1070)(3904,1064)(3887,1057)
	(3867,1048)(3844,1038)(3818,1027)
	(3789,1015)(3757,1002)(3723,988)
	(3687,974)(3649,959)(3609,944)
	(3567,929)(3524,913)(3479,898)
	(3433,883)(3386,868)(3338,853)
	(3288,839)(3237,825)(3184,812)
	(3129,799)(3072,787)(3013,776)
	(2952,766)(2888,756)(2822,747)
	(2753,740)(2682,734)(2608,729)
	(2533,726)(2457,724)(2377,724)
	(2299,727)(2223,731)(2150,737)
	(2080,744)(2013,752)(1950,762)
	(1889,772)(1831,784)(1776,796)
	(1723,809)(1672,822)(1623,836)
	(1575,851)(1529,866)(1485,881)
	(1442,896)(1401,912)(1361,928)
	(1324,943)(1288,958)(1254,972)
	(1222,986)(1194,999)(1167,1011)
	(1144,1022)(1124,1031)(1107,1040)
	(1093,1046)(1082,1052)(1073,1056)(1061,1062)
\blacken\thicklines
\path(1198.518,1035.167)(1061.000,1062.000)(1164.977,968.085)(1198.518,1035.167)
\thinlines
\path(11810,1032)(11809,1032)(11808,1031)
	(11805,1030)(11801,1029)(11795,1027)
	(11786,1024)(11775,1021)(11761,1016)
	(11744,1011)(11725,1005)(11702,998)
	(11676,990)(11648,981)(11616,972)
	(11581,961)(11543,950)(11503,938)
	(11459,926)(11414,913)(11366,899)
	(11316,885)(11264,871)(11211,857)
	(11155,842)(11099,827)(11040,812)
	(10981,798)(10920,783)(10858,769)
	(10795,754)(10730,740)(10664,726)
	(10597,713)(10529,700)(10459,687)
	(10387,675)(10314,663)(10239,651)
	(10162,640)(10083,630)(10002,620)
	(9919,611)(9833,602)(9745,594)
	(9655,587)(9562,581)(9467,576)
	(9370,572)(9271,569)(9172,567)
	(9072,567)(8964,568)(8858,571)
	(8754,575)(8653,580)(8554,587)
	(8459,594)(8368,603)(8279,612)
	(8194,623)(8113,634)(8034,645)
	(7958,658)(7884,671)(7813,684)
	(7744,698)(7678,712)(7613,727)
	(7550,742)(7489,758)(7429,774)
	(7371,790)(7315,806)(7260,822)
	(7206,838)(7154,855)(7104,871)
	(7056,887)(7010,902)(6966,918)
	(6924,932)(6885,946)(6848,960)
	(6814,972)(6782,984)(6754,995)
	(6728,1005)(6706,1013)(6686,1021)
	(6669,1027)(6655,1033)(6644,1037)
	(6636,1041)(6629,1043)(6620,1047)
\blacken\thicklines
\path(6758.595,1026.439)(6620.000,1047.000)(6728.134,957.903)(6758.595,1026.439)
\put(3568,1646){\makebox(0,0)[lb]{\smash{{\SetFigFont{9}{10.8}{\familydefault}{\mddefault}{\updefault}$a$}}}}
\put(1301,1196){\makebox(0,0)[lb]{\smash{{\SetFigFont{9}{10.8}{\familydefault}{\mddefault}{\updefault}$b$}}}}
\put(6319,1245){\makebox(0,0)[lb]{\smash{{\SetFigFont{9}{10.8}{\rmdefault}{\mddefault}{\updefault}$0$}}}}
\put(7377,1252){\makebox(0,0)[lb]{\smash{{\SetFigFont{9}{10.8}{\rmdefault}{\mddefault}{\updefault}$1$}}}}
\put(8486,1245){\makebox(0,0)[lb]{\smash{{\SetFigFont{9}{10.8}{\rmdefault}{\mddefault}{\updefault}$2$}}}}
\put(9589,1253){\makebox(0,0)[lb]{\smash{{\SetFigFont{9}{10.8}{\rmdefault}{\mddefault}{\updefault}$3$}}}}
\put(10706,1253){\makebox(0,0)[lb]{\smash{{\SetFigFont{9}{10.8}{\rmdefault}{\mddefault}{\updefault}$4$}}}}
\put(6732,1690){\makebox(0,0)[lb]{\smash{{\SetFigFont{9}{10.8}{\familydefault}{\mddefault}{\updefault}$a,b$}}}}
\put(10706,2055){\makebox(0,0)[lb]{\smash{{\SetFigFont{9}{10.8}{\familydefault}{\mddefault}{\updefault}$b$}}}}
\put(7880,1679){\makebox(0,0)[lb]{\smash{{\SetFigFont{9}{10.8}{\familydefault}{\mddefault}{\updefault}$a$}}}}
\put(9003,1678){\makebox(0,0)[lb]{\smash{{\SetFigFont{9}{10.8}{\familydefault}{\mddefault}{\updefault}$a$}}}}
\put(10122,1678){\makebox(0,0)[lb]{\smash{{\SetFigFont{9}{10.8}{\familydefault}{\mddefault}{\updefault}$a$}}}}
\put(6892,1205){\makebox(0,0)[lb]{\smash{{\SetFigFont{9}{10.8}{\familydefault}{\mddefault}{\updefault}$b$}}}}
\put(4137,2045){\makebox(0,0)[lb]{\smash{{\SetFigFont{9}{10.8}{\familydefault}{\mddefault}{\updefault}$b$}}}}
\put(3020,2052){\makebox(0,0)[lb]{\smash{{\SetFigFont{9}{10.8}{\familydefault}{\mddefault}{\updefault}$b$}}}}
\put(2352,1639){\makebox(0,0)[lb]{\smash{{\SetFigFont{9}{10.8}{\familydefault}{\mddefault}{\updefault}$a$}}}}
\put(2390,814){\makebox(0,0)[lb]{\smash{{\SetFigFont{9}{10.8}{\familydefault}{\mddefault}{\updefault}$a$}}}}
\put(8466,2059){\makebox(0,0)[lb]{\smash{{\SetFigFont{9}{10.8}{\familydefault}{\mddefault}{\updefault}$b$}}}}
\put(1835,1226){\makebox(0,0)[lb]{\smash{{\SetFigFont{9}{10.8}{\rmdefault}{\mddefault}{\updefault}$1$}}}}
\put(2922,1233){\makebox(0,0)[lb]{\smash{{\SetFigFont{9}{10.8}{\rmdefault}{\mddefault}{\updefault}$2$}}}}
\put(4099,1247){\makebox(0,0)[lb]{\smash{{\SetFigFont{9}{10.8}{\rmdefault}{\mddefault}{\updefault}$3$}}}}
\put(754,1234){\makebox(0,0)[lb]{\smash{{\SetFigFont{9}{10.8}{\rmdefault}{\mddefault}{\updefault}$0$}}}}
\put(9567,2066){\makebox(0,0)[lb]{\smash{{\SetFigFont{9}{10.8}{\familydefault}{\mddefault}{\updefault}$b$}}}}
\put(1146,1653){\makebox(0,0)[lb]{\smash{{\SetFigFont{9}{10.8}{\familydefault}{\mddefault}{\updefault}$a,b$}}}}
\put(1752,154){\makebox(0,0)[lb]{\smash{{\SetFigFont{9}{10.8}{\rmdefault}{\mddefault}{\updefault}$\cD_1=\cU_4(a,b,\emp)$}}}}
\put(11816,1246){\makebox(0,0)[lb]{\smash{{\SetFigFont{9}{10.8}{\rmdefault}{\mddefault}{\updefault}$5$}}}}
\put(11239,1670){\makebox(0,0)[lb]{\smash{{\SetFigFont{9}{10.8}{\familydefault}{\mddefault}{\updefault}$a$}}}}
\put(11809,2047){\makebox(0,0)[lb]{\smash{{\SetFigFont{9}{10.8}{\familydefault}{\mddefault}{\updefault}$b$}}}}
\put(8907,694){\makebox(0,0)[lb]{\smash{{\SetFigFont{9}{10.8}{\familydefault}{\mddefault}{\updefault}$a$}}}}
\thinlines
\put(3071.500,1757.929){\arc{394.717}{2.4948}{6.9299}}
\blacken\thicklines
\path(3231.033,1779.097)(3229.000,1639.000)(3302.998,1757.977)(3231.033,1779.097)
\end{picture}
}
\end{center}
\caption{DFA's $\cD_1$ and $\cD_2$ of $U_4(a,b,\emp)$ and $U_6(a,b,\emp)$.} 
\label{fig:binarykk}
\end{figure}
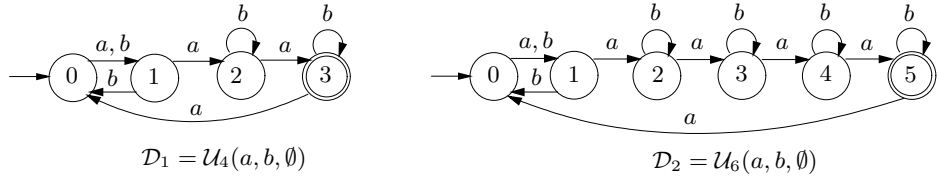

\begin{figure}[hbt]
\begin{center}
\input boolkk2.eepic
\end{center}
\caption{Direct product $\cP$ of $\cD_1=\cU_4(a,b,\emp)$ with $\cD_2=\cU_6(a,b,\emp)$.} 
\label{fig:boolkk}
\end{figure}

\begin{theorem}[$K_m\circ L_n$, $m\neq n$]
\label{thm:bool1}
For $m,n \ge 3$ and $m\neq n$, the complexity of $U_m(a,b,\emp)\circ U_n(a,b,\emp)$ is $mn$.
\end{theorem}
\begin{proof}
First it will be shown that all $mn$ states of the direct product are reachable from the initial state $(0,0)$.
Without loss of generality, assume that $m< n$.
Throughout the paper, we use the notation $q_1 \der{w} q_2$ to say that state $q_2$ is reachable from state $q_1$ by word $w$.
We have $(0,0) \der{a^m} (0,m) \der{(ab)^{n-1-m}a} (1,0)$.
For $1\le i\le m-2$, $ab$ takes $(i,0)$ to $(i+1,0)$; hence all  states in column 0 can be reached.
State $(i,j)$ can be reached from state $(i-j \pmod m,0)$ by $a^j$. 
Therefore all the states are reachable.

It remains to prove that all the states are pairwise distinguishable.
Let $H$ (for \emph{horizontal}) be the set $H=\{(m-1,0), \ldots,(m-1,n-2)\}$, 
and let $V$ (for \emph{vertical}) be  $V=\{(0,n-1),\ldots,(m-2,n-1)\}$. 
Given a state $(i,j)$, we define $d_{i,j}$ to be the minimal integer such that $a^{d_{i,j}}$ 
takes $(i,j)$ to a final state,
or infinity, if no final state is reachable by $a$'s from $(i,j)$.
This depends on the boolean operation, 
and $d_{i,j} = 0$ if and only if $(i,j)$ is final.

The boolean operations are now considered one by one.
\smallskip

\noin
{\bf Union:} The final states are those in $H \cup V \cup \{(m-1,n-1)\}$. 
We may write $d_{i,j} = \min \, \{m-1-i, n-1-j\} \le m-1$.

Let $(i,j)$ and $(k,l)$ be two distinct states, with $d_{i,j} \le d_{k,l}$. 
If $d_{i,j}<d_{k,l}$, then the two states are distinguished by $a^{d_{i,j}}$.
If $d_{i,j}=d_{k,l}=d$, apply $a^{d+1}$ to both states. 
The resulting states must be distinct and each must have at least one zero component.

If the two states are of the form $(0,n-1-g)$ and $(0,n-1-h)$, $h < g$,
then $(ab)^h$ distinguishes them. 
A symmetric argument works for $(m-1-g,0)$ and $(m-1-h,0)$.
Suppose now the states are $(0,n-1-g)$ and $(m-1-h,0)$. If $g \neq h$, then the states are distinguished by $(ab)^{\min \, \{g,h\}}$.
If $g = h$, then applying $(ab)^{g+1}$ results in the two states $(1,0)$ and $(0,1)$.
Since $d_{1,0}<d_{0,1}$ (because $m<n$), these two states are distinguished by $d_{1,0}$.
\smallskip

\noin
{\bf Symmetric Difference:} The final states are those in $H\cup V$. 

The removal of $(m-1,n-1)$ from the set of final states causes all of the $d_{i,j}$ to increase by $m$ when $m-i = n-j$, and leaves the rest unchanged.
Since all of the other $d_{i,j}$ are at most $m-1$, 
and the change maps distinct $d_{i,j}$ to distinct $d'_{i,j}$, the same argument for unequal $d_{i,j}$ applies to all pairs involving at least one of the states affected by the change. 
Since state $(m-1,n-1)$  was never used to distinguish equal $d_{i,j}$ cases in union, all remaining equality cases can be dealt with in the same way as in union.
\smallskip

\noin
{\bf Difference:} The final states are those in $H$. 

In this case only, we do not assume $m < n$. 
The $d_{i,j}$ here are as follows:
$d_{i,j} = m-1-i$ if $m-i \neq n - j$, and otherwise $d_{i,j} = 2m-1-i$. 
The same distinguishability argument applies when $d_{i,j} \neq d_{k,l}$.
Suppose $d_{i,j} = d_{k,l}$. 
Then $i = k$, and hence $j \neq l$. 
Apply $a^{m-i}$ to get two distinct states $(0,g)$ and $(0,h)$, $g \neq 0$. 
As repeated applications of $ab$ cycle through states $(0,1), (0,2), \dots, (0,n-1)$, there exists a $d$ such that $(ab)^d$ sends $(0,g)$ to $(0,n-m)$, and $(0,h)$ to a different state. 
Therefore applying $(ab)^d a^{m-1}$ maps $(0,g)$ to a non-final state, and $(0,h)$ to a final state.
\smallskip
\goodbreak

\noin
{\bf Intersection:} The only final state is $(m-1,n-1)$. 

We assume that $m < n$. If $\gcd(m,n) = 1$, then by the Chinese Remainder Theorem there is a bijection between the integers $\{0,1,\dots, mn-1\}$ and the states of the direct product given by $k \leftrightarrow (k \pmod m, k \pmod n)$. 
Applying $a$ to the state corresponding to $k$ results in the state corresponding to $k+1$. 
Thus, for state $(i,j)$ corresponding to $k$, $d_{i,j} = mn-1-k$; hence all states are distinguishable by multiple applications of $a$.

Now suppose $\gcd(m,n) > 1$. 
The states which can reach $(m-1,n-1)$ through multiple applications of $a$ are exactly those which can be written in the form $(k \pmod m, k \pmod n)$ for some integer $k$. Let $S$ denote the set of these states. Any two states in $S$ have different finite values of $d_{i,j}$, and hence are distinguishable.

Let $(i,j), (k,l) \notin S$; that is, $d_{i,j} = d_{k,l} = \infty$. 
These states can be distinguished from states in $S$ using only $a$'s.
Suppose $i \neq k$.
Apply $a^{m-i}$ to get two distinct states $(0,j')$ and $(k',l')$, $k' \neq 0$. 
Since $(0,j') \notin S$, $j' \neq 0$.
As $m < n$ and $(0,m) \in S$, there exists a $d$ such applying $(ab)^d$ to $(0,j')$ results in $(0,m)$.
Then let $d$ be the minimal integer such that applying $(ab)^d$ to the two states results in at least one state in $S$.
Because the two resulting states are distinct, they must be distinguishable.
\qed
\end{proof}

\section{Boolean Operations with One Reversed Argument}
\label{sec:onerev}
Gao and Yu~\cite{GY_FI12} studied the complexities of $K_m\cup L_n^R$ and $K_m\cap L_n^R$, and showed that they are both $m2^n-(m-1)$, with quaternary witnesses. 
These results  can be improved and extended as follows: (1)  \emph{ternary alphabets} suffice, (2)
 the \emph{same language stream} can be used for $K_m$ and $L_n$ for both union and intersection,  (3) 
 the same language stream is also a witness for two \emph{difference} operations and \emph{symmetric difference}, and 
(4) the bound for symmetric difference is $m2^n$.

The reverse $\cN^\rev$ of an NFA $\cN$ is obtained by interchanging the sets of initial and final states and reversing all transitions.

Let $\cD_1 = (Q_1, \Sigma, \delta_1, 0, \{m-1\})= \cU_m(a,b,c)$  and $\cD_2 =(Q_2, \Sigma, \delta_2, 0, \{n-1\}) = \cU_n(a,b,c)$, where $Q_1=\{0,\ldots,m-1\}$ and $Q_2=\{0,\ldots,n-1\}$.
Let $\cN_2$ be the NFA obtained by reversing $\cD_2$ and  let 
$\cR_2$ be the DFA obtained from $\cN_2$ by the subset construction.
Since the reverse of $\cN_2$ is deterministic, the subset construction applied to $\cN_2$ results in a minimal DFA, by a theorem from~\cite{Brz63}. 
Let $\cP$ be the direct product of $\cD_1$ and $\cR_2$. 
The states of $\cP$ are of the form $(i,S)$, where $S \subseteq Q_2$. 
The problem is illustrated in Fig.~\ref{fig:KcircLr}, where DFA $\cD_1$ has $m=4$ and NFA $\cN_2=\cD_2^\rev$ has $n=5$.

\begin{figure}[hbt]
\begin{center}
\setlength{\unitlength}{0.00039370in}
\begingroup\makeatletter\ifx\SetFigFont\undefined%
\gdef\SetFigFont#1#2#3#4#5{%
  \reset@font\fontsize{#1}{#2pt}%
  \fontfamily{#3}\fontseries{#4}\fontshape{#5}%
  \selectfont}%
\fi\endgroup%
{\renewcommand{\dashlinestretch}{30}
\begin{picture}(11081,2299)(0,-10)
\put(6717,109){\makebox(0,0)[lb]{\smash{{\SetFigFont{9}{10.8}{\rmdefault}{\mddefault}{\updefault}$\cN_2=\cD_2^\rev=(\cU_5(a,b,c))^\rev$}}}}
\put(1894.500,1682.929){\arc{394.717}{2.4948}{6.9299}}
\blacken\thicklines
\path(2054.033,1704.097)(2052.000,1564.000)(2125.998,1682.977)(2054.033,1704.097)
\thinlines
\put(4181.500,1690.929){\arc{394.717}{2.4948}{6.9299}}
\blacken\thicklines
\path(4341.033,1712.097)(4339.000,1572.000)(4412.998,1690.977)(4341.033,1712.097)
\thinlines
\put(806.500,1675.929){\arc{394.717}{2.4948}{6.9299}}
\blacken\thicklines
\path(966.033,1697.097)(964.000,1557.000)(1037.998,1675.977)(966.033,1697.097)
\thinlines
\put(7429.500,1698.929){\arc{394.717}{2.4948}{6.9299}}
\blacken\thicklines
\path(7589.033,1720.097)(7587.000,1580.000)(7660.998,1698.977)(7589.033,1720.097)
\thinlines
\put(6380.500,1683.929){\arc{394.717}{2.4948}{6.9299}}
\blacken\thicklines
\path(6540.033,1705.097)(6538.000,1565.000)(6611.998,1683.977)(6540.033,1705.097)
\thinlines
\put(8524.500,1690.929){\arc{394.717}{2.4948}{6.9299}}
\blacken\thicklines
\path(8684.033,1712.097)(8682.000,1572.000)(8755.998,1690.977)(8684.033,1712.097)
\thinlines
\put(9642.500,1705.929){\arc{394.717}{2.4948}{6.9299}}
\blacken\thicklines
\path(9802.033,1727.097)(9800.000,1587.000)(9873.998,1705.977)(9802.033,1727.097)
\thinlines
\put(10767.500,1690.929){\arc{394.717}{2.4948}{6.9299}}
\blacken\thicklines
\path(10927.033,1712.097)(10925.000,1572.000)(10998.998,1690.977)(10927.033,1712.097)
\thinlines
\put(817,1255){\ellipse{630}{630}}
\put(4164,1274){\ellipse{630}{630}}
\put(3035,1286){\ellipse{630}{630}}
\put(1900,1250){\ellipse{630}{630}}
\put(8522,1280){\ellipse{630}{630}}
\put(9643,1287){\ellipse{630}{630}}
\put(7429,1290){\ellipse{630}{630}}
\put(4168,1274){\ellipse{540}{540}}
\put(10758,1277){\ellipse{540}{540}}
\put(10758,1277){\ellipse{630}{630}}
\put(6372,1272){\ellipse{630}{630}}
\path(2142,1474)(2772,1474)
\blacken\thicklines
\path(2637.000,1436.500)(2772.000,1474.000)(2637.000,1511.500)(2637.000,1436.500)
\thinlines
\path(3298,1489)(3928,1489)
\blacken\thicklines
\path(3793.000,1451.500)(3928.000,1489.000)(3793.000,1526.500)(3793.000,1451.500)
\thinlines
\path(1017,1481)(1647,1481)
\blacken\thicklines
\path(1512.000,1443.500)(1647.000,1481.000)(1512.000,1518.500)(1512.000,1443.500)
\thinlines
\path(1655,1069)(1070,1069)
\blacken\thicklines
\path(1205.000,1106.500)(1070.000,1069.000)(1205.000,1031.500)(1205.000,1106.500)
\thinlines
\path(12,1264)(499,1264)
\blacken\thicklines
\path(379.000,1234.000)(499.000,1264.000)(379.000,1294.000)(379.000,1234.000)
\blacken\path(8136.000,1459.500)(8271.000,1497.000)(8136.000,1534.500)(8136.000,1459.500)
\thinlines
\path(8271,1497)(7686,1497)
\path(6597,1497)(7182,1497)
\blacken\thicklines
\path(7047.000,1459.500)(7182.000,1497.000)(7047.000,1534.500)(7047.000,1459.500)
\blacken\path(9239.000,1459.500)(9374.000,1497.000)(9239.000,1534.500)(9239.000,1459.500)
\thinlines
\path(9374,1497)(8789,1497)
\blacken\thicklines
\path(10363.000,1459.500)(10498.000,1497.000)(10363.000,1534.500)(10363.000,1459.500)
\thinlines
\path(10498,1497)(9913,1497)
\blacken\thicklines
\path(10047.000,1151.500)(9912.000,1114.000)(10047.000,1076.500)(10047.000,1151.500)
\thinlines
\path(9912,1114)(10497,1114)
\path(5548,1279)(6035,1279)
\blacken\thicklines
\path(5915.000,1249.000)(6035.000,1279.000)(5915.000,1309.000)(5915.000,1249.000)
\thinlines
\path(3919,1039)(3918,1039)(3916,1038)
	(3913,1036)(3907,1034)(3899,1030)
	(3888,1025)(3874,1019)(3857,1012)
	(3837,1003)(3814,993)(3788,982)
	(3759,970)(3727,957)(3693,943)
	(3657,929)(3619,914)(3579,899)
	(3537,884)(3494,868)(3449,853)
	(3403,838)(3356,823)(3308,808)
	(3258,794)(3207,780)(3154,767)
	(3099,754)(3042,742)(2983,731)
	(2922,721)(2858,711)(2792,702)
	(2723,695)(2652,689)(2578,684)
	(2503,681)(2427,679)(2347,679)
	(2269,682)(2193,686)(2120,692)
	(2050,699)(1983,707)(1920,717)
	(1859,727)(1801,739)(1746,751)
	(1693,764)(1642,777)(1593,791)
	(1545,806)(1499,821)(1455,836)
	(1412,851)(1371,867)(1331,883)
	(1294,898)(1258,913)(1224,927)
	(1192,941)(1164,954)(1137,966)
	(1114,977)(1094,986)(1077,995)
	(1063,1001)(1052,1007)(1043,1011)(1031,1017)
\blacken\thicklines
\path(1168.518,990.167)(1031.000,1017.000)(1134.977,923.085)(1168.518,990.167)
\thinlines
\path(10625,995)(10624,995)(10623,994)
	(10620,993)(10615,992)(10609,989)
	(10600,986)(10588,982)(10574,977)
	(10557,972)(10537,965)(10515,957)
	(10489,949)(10460,939)(10429,929)
	(10395,918)(10358,906)(10320,894)
	(10279,881)(10236,869)(10191,855)
	(10145,842)(10097,828)(10047,815)
	(9997,801)(9945,788)(9892,774)
	(9837,761)(9781,748)(9724,736)
	(9665,723)(9605,712)(9543,700)
	(9480,689)(9414,679)(9346,669)
	(9276,659)(9204,651)(9129,642)
	(9051,635)(8971,629)(8889,623)
	(8804,619)(8717,615)(8629,613)
	(8540,612)(8447,613)(8356,615)
	(8266,618)(8178,623)(8094,628)
	(8012,635)(7933,643)(7858,651)
	(7785,660)(7714,670)(7647,680)
	(7581,691)(7518,703)(7457,715)
	(7398,727)(7340,740)(7284,753)
	(7229,767)(7176,780)(7125,794)
	(7074,808)(7025,822)(6978,836)
	(6933,850)(6889,864)(6847,877)
	(6807,890)(6770,902)(6735,914)
	(6702,925)(6673,936)(6646,945)
	(6622,953)(6601,961)(6583,967)
	(6569,973)(6557,977)(6547,981)
	(6540,983)(6530,987)
\blacken\thicklines
\path(6669.272,971.680)(6530.000,987.000)(6641.417,902.044)(6669.272,971.680)
\put(1826,1975){\makebox(0,0)[lb]{\smash{{\SetFigFont{9}{10.8}{\familydefault}{\mddefault}{\updefault}$c$}}}}
\put(2292,769){\makebox(0,0)[lb]{\smash{{\SetFigFont{9}{10.8}{\familydefault}{\mddefault}{\updefault}$a,c$}}}}
\put(3538,1601){\makebox(0,0)[lb]{\smash{{\SetFigFont{9}{10.8}{\familydefault}{\mddefault}{\updefault}$a$}}}}
\put(1271,1151){\makebox(0,0)[lb]{\smash{{\SetFigFont{9}{10.8}{\familydefault}{\mddefault}{\updefault}$b$}}}}
\put(3951,1968){\makebox(0,0)[lb]{\smash{{\SetFigFont{9}{10.8}{\familydefault}{\mddefault}{\updefault}$b$}}}}
\put(688,1961){\makebox(0,0)[lb]{\smash{{\SetFigFont{9}{10.8}{\familydefault}{\mddefault}{\updefault}$c$}}}}
\put(2811,1998){\makebox(0,0)[lb]{\smash{{\SetFigFont{9}{10.8}{\familydefault}{\mddefault}{\updefault}$b,c$}}}}
\put(1116,1593){\makebox(0,0)[lb]{\smash{{\SetFigFont{9}{10.8}{\familydefault}{\mddefault}{\updefault}$a,b$}}}}
\put(2367,1594){\makebox(0,0)[lb]{\smash{{\SetFigFont{9}{10.8}{\familydefault}{\mddefault}{\updefault}$a$}}}}
\put(8367,702){\makebox(0,0)[lb]{\smash{{\SetFigFont{9}{10.8}{\familydefault}{\mddefault}{\updefault}$a,c$}}}}
\put(6289,1200){\makebox(0,0)[lb]{\smash{{\SetFigFont{9}{10.8}{\rmdefault}{\mddefault}{\updefault}$4$}}}}
\put(7347,1207){\makebox(0,0)[lb]{\smash{{\SetFigFont{9}{10.8}{\rmdefault}{\mddefault}{\updefault}$3$}}}}
\put(8456,1200){\makebox(0,0)[lb]{\smash{{\SetFigFont{9}{10.8}{\rmdefault}{\mddefault}{\updefault}$2$}}}}
\put(9559,1208){\makebox(0,0)[lb]{\smash{{\SetFigFont{9}{10.8}{\rmdefault}{\mddefault}{\updefault}$1$}}}}
\put(10676,1208){\makebox(0,0)[lb]{\smash{{\SetFigFont{9}{10.8}{\rmdefault}{\mddefault}{\updefault}$0$}}}}
\put(8973,1633){\makebox(0,0)[lb]{\smash{{\SetFigFont{9}{10.8}{\familydefault}{\mddefault}{\updefault}$a$}}}}
\put(9987,1622){\makebox(0,0)[lb]{\smash{{\SetFigFont{9}{10.8}{\familydefault}{\mddefault}{\updefault}$a,b$}}}}
\put(10624,1995){\makebox(0,0)[lb]{\smash{{\SetFigFont{9}{10.8}{\familydefault}{\mddefault}{\updefault}$c$}}}}
\put(9477,2006){\makebox(0,0)[lb]{\smash{{\SetFigFont{9}{10.8}{\familydefault}{\mddefault}{\updefault}$c$}}}}
\put(8316,2029){\makebox(0,0)[lb]{\smash{{\SetFigFont{9}{10.8}{\familydefault}{\mddefault}{\updefault}$b,c$}}}}
\put(7197,2015){\makebox(0,0)[lb]{\smash{{\SetFigFont{9}{10.8}{\familydefault}{\mddefault}{\updefault}$b,c$}}}}
\put(6303,2017){\makebox(0,0)[lb]{\smash{{\SetFigFont{9}{10.8}{\familydefault}{\mddefault}{\updefault}$b$}}}}
\put(6776,1640){\makebox(0,0)[lb]{\smash{{\SetFigFont{9}{10.8}{\familydefault}{\mddefault}{\updefault}$a$}}}}
\put(7857,1634){\makebox(0,0)[lb]{\smash{{\SetFigFont{9}{10.8}{\familydefault}{\mddefault}{\updefault}$a$}}}}
\put(10109,1205){\makebox(0,0)[lb]{\smash{{\SetFigFont{9}{10.8}{\familydefault}{\mddefault}{\updefault}$b$}}}}
\put(1309,116){\makebox(0,0)[lb]{\smash{{\SetFigFont{9}{10.8}{\rmdefault}{\mddefault}{\updefault}$\cD_1=\cU_4(a,b,c)$}}}}
\put(716,1167){\makebox(0,0)[lb]{\smash{{\SetFigFont{9}{10.8}{\rmdefault}{\mddefault}{\updefault}$0$}}}}
\put(1790,1166){\makebox(0,0)[lb]{\smash{{\SetFigFont{9}{10.8}{\rmdefault}{\mddefault}{\updefault}$1$}}}}
\put(2922,1173){\makebox(0,0)[lb]{\smash{{\SetFigFont{9}{10.8}{\rmdefault}{\mddefault}{\updefault}$2$}}}}
\put(4084,1187){\makebox(0,0)[lb]{\smash{{\SetFigFont{9}{10.8}{\rmdefault}{\mddefault}{\updefault}$3$}}}}
\thinlines
\put(3041.500,1712.929){\arc{394.717}{2.4948}{6.9299}}
\blacken\thicklines
\path(3201.033,1734.097)(3199.000,1594.000)(3272.998,1712.977)(3201.033,1734.097)
\end{picture}
}
\end{center}
\caption{DFA $\cD_1=\cU_4(a,b,c)$ and NFA $\cN_2=\cD_2^\rev=(\cU_5(a,b,c))^\rev$.} 
\label{fig:KcircLr}
\end{figure}

First we derive upper bounds for two differences and for symmetric difference.
\begin{proposition}
\label{prop:oneBound}
Let $K_m$ and $L_n$ be two regular languages with complexities $m$ and $n$.
Then the complexities of $K_m \bs L_n^R$ and $L_n^R \bs K_m$ are at most $m2^n - (m-1)$,
and that of $K_m \oplus L_n^R$ is at most $m2^n$.
\end{proposition}
\begin{proof}
Let $\cD_1 = (Q_1, \Sigma, \delta_1, q_1, F_1)$ and $\cD_2 = (Q_2, \Sigma, \delta_2, q_2, F_2)$ be the minimal DFA's of $K_m$ and $L_n$.
Consider the direct product $\cP$ of $\cD_1$ and $\cR_2$, which is the determinized version of $\cD_2^\rev$. 
With appropriate assignments of final states, $\cP$ can accept the languages $K_m \bs L_n^R$, $L_n^R \bs K_m$, and $K_m \oplus L_n^R$.
The states of $\cP$ are of the form $(i, S)$ where $i \in Q_1$ and $S \subseteq Q_2$.
Therefore there are at most $m2^n$ states in $\cP$, thus proving the bound for $K_m \oplus L_n^R$.

Note that any state of the form $(i, \emp)$ is mapped to a state of the same form under any input $x \in \Sigma$.
Also, any state of the form $(i, Q_2)$ is mapped to a state of the same form since $\cD_2$ is complete.
For $K_m \bs L_n^R$, all $m$ states of the form $(i, Q_2)$ are non-final, and thus indistinguishable.
For $L_n^R \bs K_m$, all $m$ states of the form $(i, \emp)$ are non-final and indistinguishable.
Therefore  $\cP$ contains at most $m2^n - (m-1)$ distinguishable states for $K_m \bs L_n^R$ and $L_n^R \bs K_m$.
\qed
\end{proof}

\begin{theorem}[$K\circ L^R$]
\label{thm:KRunionL}
For $m,n \ge 3$,   the complexities of the four languages 
$U_m(a,b,c)\cup (U_n(a,b,c))^R$, 
$U_m(a,b,c)\cap (U_n(a,b,c))^R$, 
$U_m(a,b,c)\setminus (U_n(a,b,c))^R$, \\ and 
 $(U_n(a,b,c))^R\setminus U_m(a,b,c)$
are all $m2^n-(m-1)$, and that of $U_m(a,b,c)\oplus (U_n(a,b,c))^R$ is 
$m2^n$.
\end{theorem}
\begin{proof}
Let $\cK_m=\cU_m(a,b,c)$ and $\cL_n=\cU_n(a,b,c)$;
the various related automata are defined as above.
It is  known from~\cite{BrTa12} that the complexity of  $L_n^R$ is $2^n$; hence that of $K_m\circ L_n^R$ is at most $m2^n$.
We first show that all $m2^n$ states of $\cP$ are reachable.

The initial state is $(0,\{n-1\})$.
We have $(0, \{n-1\}) \der{c} (0,\emp) \der{a^i} (i,\emp)$ for $i=1,\ldots,n-1$. 
Input $ab$ acts on $\cN_2$ as the cycle $(n-1,n-2,\dots, 2, 0)$ and sends $0$ to $0$ in $\cD_1$. 
Therefore all states of the form $(0, \{j\})$ with $j \neq 1$ are reachable from $(0, \{n-1\})$ by repeated applications of $ab$. 
If $n \nmid m$, then $\{1 + m \pmod n \} \neq \{1\}$ and $(0, \{1\})$ is reachable by $a^m$ from $(0,\{1+m \pmod n\})$. 
If $n \mid m$, then $m-1 \equiv n-1 \pmod n$; so we have $(0, \{0\}) \der{a^{m-1}} ({m-1}, \{1\}) \der{c} (0, \{1\})$.
For $i=1,\ldots,m-1$, $(i,\{j\})$ is reached from $({0}, \{i+j \pmod n\})$ by $a^i$.
So all  states of the form $(i,S)$, where $|S| \le 1$, are reachable.

Now suppose it is possible to reach all states of the form $(i,S)$, where
$|S| = k$, $k \ge 1$. 
We want to show it is possible to reach all states $(i, S)$ with $|S| = k + 1$. 
The transformations $a$ and $b$ generate all permutations of states in $\cN_2$. 
Since $|S| \ge 2$, there is a word $w \in \{a,b\}^*$ and $S' \subseteq Q_2$ of size $k+1$ with $0, n-1 \in S'$ such that $S' \der{w} S$. 
Moreover, $w$ also causes a permutation of the states in $\cD_1$. 
Therefore it suffices to show the reachability of all states of the form $(i, S)$, where $|S| = k + 1$ and $0, n-1 \in S$.

Let $S \subseteq Q_2$, $|S| = k+1$, and $0, n-1 \in S$.
Define $S' = S \bs \{n-1\}$. 
All states of the form $(i, S')$ are reachable, and 
$(i, S') \der{c} (i, S)$ for all $i \le m-2$.
For  state $({m-1}, S)$ there are three cases:
\be
\item $m \nmid n$. 
State ${m-1-n} \pmod m$ is distinct from ${m-1}$. Therefore we have 
$({m-1-n}, S) \der{a^n} ({m-1}, S)$.

\item $m = n = 3$.
Note that $a^2ba$ is a transposition $(1,2)$ in $\cD_1$ and $(0,2)$ in $\cN_2$.
Thus  $(1, S) \der{a^2ba} (2,S)$, since $0,2\in S$.

\item $m \mid n$, $n \ge 4$. 
Define $S''$ to be the result of applying the transposition $a^2 b a^{n-2} : (2,3)$ 
in $\cN_2$ to $S'$. 
So $S''$ is like $S'$ with 2 and 3 transposed, if present. 
Since $S'$ is $S$ without $n-1$, and we have $0\in S$, we also have  $0\in S'$ and $0\in S''$.
Applying $c$ to $S''$ adds $n-1$.
Applying $c a^2 ba^{n-2}$ to $S''$  adds $n-1$ and transposes 2 and 3, if present; hence the result is $S$. 
Since $m \mid n$, $a^{n-2}$ is the same transformation as $a^{m-2}$ in $\cD_1$; 
hence $a^2 ba^{n-2}$ is the transposition $({m-2}, {m-1})$ in $\cD_1$.
It follows that $({m-2}, S'') \der{ca^2 ba^{n-2}} (m-1, S)$.
\ee

Therefore all $m2^n$ states are reachable, and it remains to find the number of pairwise indistinguishable states for each operation.

We claim that if $S,T \subseteq Q_2$ are distinct states of $\cR_2$, then there is an input which takes this pair of states to $\emp$ and $Q_2$. 
First suppose $0 \in S \bs T$. 
Then applying $c$ results in two states $S_1$ and $T_1$ such that $0, n-1 \in S_1 \bs T_1$. 
For $k \ge 2$, define $S_k$ and $T_k$ as the states obtained by applying $a^{n-1}c$ to $S_{k-1}$ and $T_{k-1}$, respectively. 
Then $0,1,\dots, k-1, n-1  \in S_k \bs T_k$. 
It follows that $S_{n-1} = Q_2$ and $T_{n-1} = \emp$.
In general, if $j \in S \bs T$, then applying $a^j$ sends $S$ and $T$ to the case $0 \in S \bs T$, and so the claim is true.

Sets $Q_2$ and $\emp$ are mapped to themselves under all inputs $x \in \Sigma$.
Also, $Q_2$ is final and $\emp$ non-final in $\cR_2$. 
Therefore any states of the form $(i, Q_2)$ and $(j, \emp)$ are distinguishable for the boolean operations as follows:
\bi
\item $K_m \cup L_n^R$, $L_n^R \bs K_m$, and $K_m \oplus L_n^R$: apply $a^k$, $k \notin \{m-1-i, m-1-j\}$, to send $i$ and $j$ to non-final states.
\item $K_m \cap L_n^R$: apply $a^{m-1-i}$ so that $i$ gets mapped to a final state.
\item $K_m \bs L_n^R$: apply $a^{m-1-j}$ so that $j$ gets mapped to a final state.
\ei

Thus any two states $(i, S)$ and $(j,T)$ with $S \neq T$ are distinguishable for all five boolean operations.
Now consider states of the form $(i, S)$ and $(j, S)$, $i < j$. 

{\bf Case 1:} $S = \emp$. Since all states of the form $(i, \emp)$ are non-final for $K_m \cap L_n^R$ and $L_n^R \bs K_m$, these states are indistinguishable. For the other three boolean operations, apply $a^{m-1-j}$ to get the distinguishable states $(k, \emp)$, $({m-1}, \emp)$, $k \neq m-1$. 

{\bf Case 2:} $S \neq \emp$, $S$ is non-final (i.e., $0 \notin S$). 
In $\cD_1$, $ba$ causes the cycle $(0,2,3,\dots, {m-1})$, 
and in $\cN_2$, $ba : (n-1, n-2,\dots, 1)$. 
Since $i \neq j$, at least one of $i$ and $j$ is not equal to 1. 
Therefore we can apply $(ba)^d$ for some $d$ so that the states become $({m-1}, S')$, $(k, S')$ where $S'$ is non-final, and $k \neq m-1$. 
This distinguishes the states for $K_m \cup L_n^R$, $K_m \oplus L_n^R$, and $K_m \bs L_n^R$. 
For the other two operations, apply a cyclic shift $a^r$ so that $S$ is mapped to some $S''$ and $0 \in S''$, and the pair of states is now in Case 3.

{\bf Case 3:} $S \neq Q_n$, $0 \in S$. Again, apply $(ba)^p$ for some $p$ so that the states become $({m-1}, S')$, $(k, S')$, 
$S'$ is final, and $k \neq m-1$.
This distinguishes the states for $K_m \cap L_n^R$ and $L_n^R \bs K_m$.
For the other three operations, apply a cyclic shift $a^r$ so that $S$ is mapped to $S''$, and $0 \notin S''$, so that Case 2 now applies.

{\bf Case 4:} $S = Q_n$. Since all states of the form $(i, Q_2)$ are final for $K_m \cup L_n^R$ and non-final for $K_m \bs L_n^R$, the states are indistinguishable for these cases. For the other three boolean operations, apply $a^{m-1-j}$ to get the states $(k, Q_2)$, $({m-1}, Q_2)$, $k \neq m-1$. This distinguishes the states.

Therefore for symmetric difference, all $m2^n$ states are distinguishable. 
For the other four operations, exactly $m$ states are equivalent, thus proving the bounds in the theorem.
\qed
\end{proof}

\section{Boolean Operations with Two Reversed Arguments}
\label{sec:tworev}
 Note that $(K\circ L)^R=K^R \circ L^R$ for all four boolean operations.
 Liu, Martin-Vide, A.~Salomaa, and Yu~\cite{LMSY_IC08} showed that $(2^m-1)(2^n-1) + 1$ is a tight upper bound for  $K^R\cup L^R$ and
 $K^R\cap L^R$, and that the bound is met by ternary witnesses. 
We first derive upper bounds for difference and symmetric difference.
\begin{proposition}
\label{prop:2rev_bound}
Let $K_m$ and $L_n$ be two regular languages with complexities $m$ and $n$.
Then the complexity of $K_m^R \bs L_n^R$ is at most $(2^m-1)(2^n-1) + 1$,
and the complexity of $K_m^R \oplus L_n^R$ is at most $2^{m+n-1}$.
\end{proposition}
\begin{proof}
Let $\cD_1 = (Q_1, \Sigma, \delta_1, q_1, F_1)$ and $\cD_2 = (Q_2, \Sigma, \delta_2, q_2, F_2)$ be the minimal DFA's of $K_m$ and $L_n$.
As in Proposition \ref{prop:oneBound}, we apply the standard subset construction to the NFA's $\cN_1$ and $\cN_2$ obtained by reversing $\cD_1$ and $\cD_2$, and then construct their direct product DFA $\cP$.
The states of $\cP$ are of the form $(S,T)$ where $S \subseteq Q_1$ and $T \subseteq Q_2$; hence $\cP$ has $2^{m+n}$ states.

For $K_m^R \bs L_n^R$, all states of the form $(\emp, T)$ and $(S, Q_2)$ are non-final.
Moreover, because $\cD_2$ is complete, applying any input $x \in \Sigma$ leads to a state of the same form.
Therefore these states are indistinguishable. 
As there are $(2^m-1)(2^n-1)$ states \emph{not} of this form, $\cP$ has at most $(2^m-1)(2^n-1) + 1$ distinguishable states.

For $K_m^R \oplus L_n^R$, we note that $(S,T)$ is final if and only if $(\bar S, \bar T)$ is final, where $\bar S=Q_1\setminus S$ and $\bar T=Q_2\setminus T$.
Let $S \subseteq Q_1$ be a subset of states of $\cN_1$;  apply $x \in \Sigma$ to get a state $S'$.
Then $i \in S'$ if and only if $\delta_1(i,x) \in S$.
It follows that $S$ and $\bar S$ are mapped to a pair $S', \bar S'$, i.e., complementary states are mapped to complementary states in $\cN_1$ and $\cN_2$.
Therefore complementary states are indistinguishable. Since every state has exactly one complement, $\cP$ has at most $2^{m+n-1}$ distinguishable states.
\qed
\end{proof}

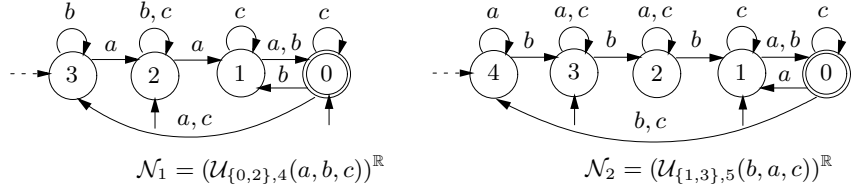
\begin{figure}[hbt]
\begin{center}
\setlength{\unitlength}{0.00039370in}
\begingroup\makeatletter\ifx\SetFigFont\undefined%
\gdef\SetFigFont#1#2#3#4#5{%
  \reset@font\fontsize{#1}{#2pt}%
  \fontfamily{#3}\fontseries{#4}\fontshape{#5}%
  \selectfont}%
\fi\endgroup%
{\renewcommand{\dashlinestretch}{30}
\begin{picture}(11098,2487)(0,-10)
\put(738,1354){\makebox(0,0)[lb]{\smash{{\SetFigFont{9}{10.8}{\rmdefault}{\mddefault}{\updefault}$3$}}}}
\put(1908.500,1869.929){\arc{394.717}{2.4948}{6.9299}}
\blacken\thicklines
\path(2068.033,1891.097)(2066.000,1751.000)(2139.998,1869.977)(2068.033,1891.097)
\thinlines
\put(4195.500,1877.929){\arc{394.717}{2.4948}{6.9299}}
\blacken\thicklines
\path(4355.033,1899.097)(4353.000,1759.000)(4426.998,1877.977)(4355.033,1899.097)
\thinlines
\put(820.500,1862.929){\arc{394.717}{2.4948}{6.9299}}
\blacken\thicklines
\path(980.033,1884.097)(978.000,1744.000)(1051.998,1862.977)(980.033,1884.097)
\thinlines
\put(7443.500,1885.929){\arc{394.717}{2.4948}{6.9299}}
\blacken\thicklines
\path(7603.033,1907.097)(7601.000,1767.000)(7674.998,1885.977)(7603.033,1907.097)
\thinlines
\put(6394.500,1870.929){\arc{394.717}{2.4948}{6.9299}}
\blacken\thicklines
\path(6554.033,1892.097)(6552.000,1752.000)(6625.998,1870.977)(6554.033,1892.097)
\thinlines
\put(8538.500,1877.929){\arc{394.717}{2.4948}{6.9299}}
\blacken\thicklines
\path(8698.033,1899.097)(8696.000,1759.000)(8769.998,1877.977)(8698.033,1899.097)
\thinlines
\put(9656.500,1892.929){\arc{394.717}{2.4948}{6.9299}}
\blacken\thicklines
\path(9816.033,1914.097)(9814.000,1774.000)(9887.998,1892.977)(9816.033,1914.097)
\thinlines
\put(10781.500,1877.929){\arc{394.717}{2.4948}{6.9299}}
\blacken\thicklines
\path(10941.033,1899.097)(10939.000,1759.000)(11012.998,1877.977)(10941.033,1899.097)
\thinlines
\put(831,1442){\ellipse{630}{630}}
\put(3049,1473){\ellipse{630}{630}}
\put(1914,1437){\ellipse{630}{630}}
\put(8536,1467){\ellipse{630}{630}}
\put(9657,1474){\ellipse{630}{630}}
\put(7443,1477){\ellipse{630}{630}}
\put(10775,1463){\ellipse{630}{630}}
\put(6386,1459){\ellipse{630}{630}}
\put(10771,1467){\ellipse{540}{540}}
\put(4183,1461){\ellipse{540}{540}}
\put(4178,1461){\ellipse{630}{630}}
\blacken\thicklines
\path(2622.000,1608.500)(2757.000,1646.000)(2622.000,1683.500)(2622.000,1608.500)
\thinlines
\path(2757,1646)(2172,1646)
\blacken\thicklines
\path(3777.000,1616.500)(3912.000,1654.000)(3777.000,1691.500)(3777.000,1616.500)
\thinlines
\path(3912,1654)(3327,1654)
\blacken\thicklines
\path(1534.000,1616.500)(1669.000,1654.000)(1534.000,1691.500)(1534.000,1616.500)
\thinlines
\path(1669,1654)(1084,1654)
\blacken\thicklines
\path(3431.000,1316.500)(3296.000,1279.000)(3431.000,1241.500)(3431.000,1316.500)
\thinlines
\path(3296,1279)(3926,1279)
\blacken\thicklines
\path(7062.000,1646.500)(7197.000,1684.000)(7062.000,1721.500)(7062.000,1646.500)
\thinlines
\path(7197,1684)(6612,1684)
\blacken\thicklines
\path(8157.000,1646.500)(8292.000,1684.000)(8157.000,1721.500)(8157.000,1646.500)
\thinlines
\path(8292,1684)(7707,1684)
\blacken\thicklines
\path(9245.000,1638.500)(9380.000,1676.000)(9245.000,1713.500)(9245.000,1638.500)
\thinlines
\path(9380,1676)(8795,1676)
\blacken\thicklines
\path(10069.000,1316.500)(9934.000,1279.000)(10069.000,1241.500)(10069.000,1316.500)
\thinlines
\path(9934,1279)(10519,1279)
\blacken\thicklines
\path(10370.000,1646.500)(10505.000,1684.000)(10370.000,1721.500)(10370.000,1646.500)
\thinlines
\path(10505,1684)(9920,1684)
\dashline{60.000}(5584,1474)(6071,1474)
\blacken\thicklines
\path(5951.000,1444.000)(6071.000,1474.000)(5951.000,1504.000)(5951.000,1444.000)
\thinlines
\path(1916,746)(1916,1106)
\blacken\thicklines
\path(1953.500,971.000)(1916.000,1106.000)(1878.500,971.000)(1953.500,971.000)
\thinlines
\path(4211,791)(4211,1151)
\blacken\thicklines
\path(4248.500,1016.000)(4211.000,1151.000)(4173.500,1016.000)(4248.500,1016.000)
\thinlines
\path(7451,791)(7451,1151)
\blacken\thicklines
\path(7488.500,1016.000)(7451.000,1151.000)(7413.500,1016.000)(7488.500,1016.000)
\thinlines
\path(9671,783)(9671,1143)
\blacken\thicklines
\path(9708.500,1008.000)(9671.000,1143.000)(9633.500,1008.000)(9708.500,1008.000)
\thinlines
\dashline{60.000}(12,1466)(499,1466)
\blacken\thicklines
\path(379.000,1436.000)(499.000,1466.000)(379.000,1496.000)(379.000,1436.000)
\thinlines
\path(4022,1153)(4021,1153)(4019,1151)
	(4016,1149)(4010,1146)(4003,1142)
	(3992,1135)(3979,1128)(3962,1118)
	(3943,1107)(3920,1094)(3895,1080)
	(3866,1064)(3835,1047)(3801,1029)
	(3765,1010)(3726,990)(3686,969)
	(3644,948)(3600,927)(3555,906)
	(3509,885)(3461,864)(3412,843)
	(3362,823)(3310,803)(3258,784)
	(3203,765)(3147,748)(3090,731)
	(3030,714)(2968,699)(2904,685)
	(2837,672)(2769,661)(2697,651)
	(2624,642)(2548,636)(2471,631)
	(2394,629)(2313,629)(2233,632)
	(2156,638)(2082,645)(2010,654)
	(1942,665)(1877,678)(1815,691)
	(1756,706)(1700,722)(1645,740)
	(1593,757)(1543,776)(1495,796)
	(1448,816)(1403,836)(1360,857)
	(1317,878)(1277,899)(1238,921)
	(1200,942)(1165,962)(1131,982)
	(1099,1001)(1070,1020)(1044,1036)
	(1020,1052)(999,1066)(980,1078)
	(965,1088)(952,1097)(942,1103)
	(935,1108)(924,1116)
\blacken\thicklines
\path(1055.236,1066.924)(924.000,1116.000)(1011.123,1006.269)(1055.236,1066.924)
\thinlines
\path(10640,1156)(10639,1156)(10638,1155)
	(10635,1153)(10630,1151)(10624,1147)
	(10615,1143)(10604,1137)(10590,1129)
	(10573,1120)(10553,1110)(10530,1098)
	(10505,1085)(10476,1071)(10445,1055)
	(10411,1038)(10375,1021)(10336,1002)
	(10295,983)(10252,963)(10208,943)
	(10161,922)(10114,901)(10064,880)
	(10014,860)(9962,839)(9908,818)
	(9854,798)(9798,779)(9740,759)
	(9681,740)(9621,722)(9558,704)
	(9494,687)(9428,671)(9359,656)
	(9289,641)(9215,627)(9139,614)
	(9061,603)(8979,592)(8895,583)
	(8809,576)(8721,570)(8631,566)
	(8540,564)(8449,564)(8359,566)
	(8271,570)(8185,576)(8101,583)
	(8020,592)(7942,602)(7866,613)
	(7793,625)(7723,638)(7655,652)
	(7589,667)(7525,683)(7463,699)
	(7403,716)(7345,733)(7288,752)
	(7232,770)(7178,789)(7125,808)
	(7074,828)(7024,848)(6975,867)
	(6928,887)(6882,907)(6838,926)
	(6795,945)(6755,964)(6717,981)
	(6681,998)(6648,1014)(6617,1029)
	(6589,1043)(6563,1056)(6541,1067)
	(6521,1077)(6505,1085)(6491,1092)
	(6480,1098)(6471,1103)(6464,1106)(6455,1111)
\blacken\thicklines
\path(6591.223,1078.219)(6455.000,1111.000)(6554.800,1012.657)(6591.223,1078.219)
\put(6303,1387){\makebox(0,0)[lb]{\smash{{\SetFigFont{9}{10.8}{\rmdefault}{\mddefault}{\updefault}$4$}}}}
\put(7361,1394){\makebox(0,0)[lb]{\smash{{\SetFigFont{9}{10.8}{\rmdefault}{\mddefault}{\updefault}$3$}}}}
\put(7864,1821){\makebox(0,0)[lb]{\smash{{\SetFigFont{9}{10.8}{\familydefault}{\mddefault}{\updefault}$b$}}}}
\put(8987,1820){\makebox(0,0)[lb]{\smash{{\SetFigFont{9}{10.8}{\familydefault}{\mddefault}{\updefault}$b$}}}}
\put(1248,1766){\makebox(0,0)[lb]{\smash{{\SetFigFont{9}{10.8}{\familydefault}{\mddefault}{\updefault}$a$}}}}
\put(2395,1774){\makebox(0,0)[lb]{\smash{{\SetFigFont{9}{10.8}{\familydefault}{\mddefault}{\updefault}$a$}}}}
\put(3394,1780){\makebox(0,0)[lb]{\smash{{\SetFigFont{9}{10.8}{\familydefault}{\mddefault}{\updefault}$a,b$}}}}
\put(3535,1361){\makebox(0,0)[lb]{\smash{{\SetFigFont{9}{10.8}{\familydefault}{\mddefault}{\updefault}$b$}}}}
\put(9979,1817){\makebox(0,0)[lb]{\smash{{\SetFigFont{9}{10.8}{\familydefault}{\mddefault}{\updefault}$a,b$}}}}
\put(6775,1812){\makebox(0,0)[lb]{\smash{{\SetFigFont{9}{10.8}{\familydefault}{\mddefault}{\updefault}$b$}}}}
\put(4070,2192){\makebox(0,0)[lb]{\smash{{\SetFigFont{9}{10.8}{\familydefault}{\mddefault}{\updefault}$c$}}}}
\put(2967,2207){\makebox(0,0)[lb]{\smash{{\SetFigFont{9}{10.8}{\familydefault}{\mddefault}{\updefault}$c$}}}}
\put(1711,2199){\makebox(0,0)[lb]{\smash{{\SetFigFont{9}{10.8}{\familydefault}{\mddefault}{\updefault}$b,c$}}}}
\put(724,2193){\makebox(0,0)[lb]{\smash{{\SetFigFont{9}{10.8}{\familydefault}{\mddefault}{\updefault}$b$}}}}
\put(6287,2204){\makebox(0,0)[lb]{\smash{{\SetFigFont{9}{10.8}{\familydefault}{\mddefault}{\updefault}$a$}}}}
\put(8283,2209){\makebox(0,0)[lb]{\smash{{\SetFigFont{9}{10.8}{\familydefault}{\mddefault}{\updefault}$a,c$}}}}
\put(9581,2209){\makebox(0,0)[lb]{\smash{{\SetFigFont{9}{10.8}{\familydefault}{\mddefault}{\updefault}$c$}}}}
\put(10655,2210){\makebox(0,0)[lb]{\smash{{\SetFigFont{9}{10.8}{\familydefault}{\mddefault}{\updefault}$c$}}}}
\put(7180,2217){\makebox(0,0)[lb]{\smash{{\SetFigFont{9}{10.8}{\familydefault}{\mddefault}{\updefault}$a,c$}}}}
\put(10135,1372){\makebox(0,0)[lb]{\smash{{\SetFigFont{9}{10.8}{\familydefault}{\mddefault}{\updefault}$a$}}}}
\put(8238,754){\makebox(0,0)[lb]{\smash{{\SetFigFont{9}{10.8}{\familydefault}{\mddefault}{\updefault}$b,c$}}}}
\put(2209,784){\makebox(0,0)[lb]{\smash{{\SetFigFont{9}{10.8}{\familydefault}{\mddefault}{\updefault}$a,c$}}}}
\put(10690,1380){\makebox(0,0)[lb]{\smash{{\SetFigFont{9}{10.8}{\rmdefault}{\mddefault}{\updefault}$0$}}}}
\put(9566,1380){\makebox(0,0)[lb]{\smash{{\SetFigFont{9}{10.8}{\rmdefault}{\mddefault}{\updefault}$1$}}}}
\put(8455,1372){\makebox(0,0)[lb]{\smash{{\SetFigFont{9}{10.8}{\rmdefault}{\mddefault}{\updefault}$2$}}}}
\put(1683,109){\makebox(0,0)[lb]{\smash{{\SetFigFont{9}{10.8}{\rmdefault}{\mddefault}{\updefault}$\cN_1=(\cU_{\{0,2\},4}(a,b,c))^\rev$}}}}
\put(7594,115){\makebox(0,0)[lb]{\smash{{\SetFigFont{9}{10.8}{\rmdefault}{\mddefault}{\updefault}$\cN_2=(\cU_{\{1,3\},5}(b,a,c))^\rev$}}}}
\put(4098,1367){\makebox(0,0)[lb]{\smash{{\SetFigFont{9}{10.8}{\rmdefault}{\mddefault}{\updefault}$0$}}}}
\put(2958,1374){\makebox(0,0)[lb]{\smash{{\SetFigFont{9}{10.8}{\rmdefault}{\mddefault}{\updefault}$1$}}}}
\put(1819,1345){\makebox(0,0)[lb]{\smash{{\SetFigFont{9}{10.8}{\rmdefault}{\mddefault}{\updefault}$2$}}}}
\thinlines
\put(3055.500,1899.929){\arc{394.717}{2.4948}{6.9299}}
\blacken\thicklines
\path(3215.033,1921.097)(3213.000,1781.000)(3286.998,1899.977)(3215.033,1921.097)
\end{picture}
}
\end{center}
\caption{NFA's $\cN_1=(\cU_{\{0,2\},4}(a,b,c))^\rev$ and $\cN_2=(\cU_{\{1,3\},5}(b,a,c))^\rev$.} 
\label{fig:KrULr}
\end{figure}

Next, we require a result concerning $\cU_{m}(a,b,c)$ and 
$\cU_{n}(b,a,c)$.
The NFA's $\cN_1=(\cU_{4}(a,b,c))^\rev$ and $\cN_2=(\cU_{5}(b,a,c))^\rev$ are shown in Fig.~\ref{fig:KrULr}, if the initial states are taken to be $3$ and $4$ as shown by the dotted arrows.

\begin{lemma}
\label{lem:2rev}
For $m,n \ge 3$, the complexities of $(U_{m}(a,b,c))^R\cup (U_{n}(b,a,c))^R$, 
$(U_{m}(a,b,c))^R\cap (U_{n}(b,a,c))^R$ and
$(U_{m}(a,b,c))^R\setminus (U_{n}(b,a,c))^R$ 
are $(2^m-1)(2^n-1) + 1$, whereas that of 
$(U_{m}(a,b,c))^R\oplus (U_{n}(b,a,c))^R$ is
 $2^{m+n-1}$, except  when $m=n=4$; then the first three complexities are 202 and the fourth is 116.
\end{lemma}
\begin{proof}
Let $\cD_1 = (Q_1, \Sigma, \delta_1, 0, \{m-1\})$ and $\cD_2 = (Q_2, \Sigma, \delta_2, 0, \{n-1\})$ be the minimal DFA's of $\cU_{m}(a,b,c)$  and $\cU_{m}(b,a,c)$.
Let  $\cN_1$  and $\cN_2$ be the NFA's obtained by reversing $\cD_1$ and $\cD_2$.
Let $\cR_1$  and $\cR_2$ be the  DFA's obtained from $\cN_1$ and $\cN_2$ by the subset construction. Since the reverses of $\cN_1$  and $\cN_2$ is deterministic, $\cR_1$  and $\cR_2$
are minimal~\cite{Brz63}.
Let $\cP$ be the direct product of $\cR_1$ and $\cR_2$.
The states of $\cP$ are of the form $(S,T)$ where $S \subseteq Q_1$ and $T \subseteq Q_2$. 

We first show that all $2^{m+n}$ states of $\cP$ are reachable if it is not the case that
$m=n=4$.
The initial state is $(\{{m-1}\}, \{n-1\})$. 
From this state, $(\emp, \emp)$ is reached by $c$. 
Also, $(\{{m-1}\}, \{n-1\}) \der{bc} (\emp, \{n-2\}) \der{b^{n-2-j}} (\emp, \{j\})$ for $j < n-2$, 
and $(\emp, \{0\}) \der{b} (\emp, \{n-1\})$.
Similarly, $(\{{m-1}\}, \{n-1\}) \der{aca^{m-2-i}} (\{i\}, \emp)$ for $i \le m-2$, and $(\{0\}, \emp) \der{a} (\{m-1\}, \emp)$.

For $i,j \ge 2$, $(\{{m-1}\}, \{n-1\}) \der{a^{m-1-i} b^{n-1-j}} (\{i\}, \{j\})$.
For the other four states, we have the transformations 
$(\{2\}, \{3\}) \der{ab^2} (\{1\}, \{1\}) \der{a} (\{0\}, \{0\})$
and 
$(\{2\}, \{2\}) \der{ab^2} (\{1\},\{0\}) \der a (\{0\}, \{1\})$.
Therefore all states of the form $(S,T)$ with $|S|, |T| \le 1$ are reachable.

Suppose all states of the form $(\{i\}, T)$ are reachable for $|T| = k$, $k \ge 1$. 
Let $T \subseteq Q_2$ with $|T| = k + 1$ and $0,n-1 \in T$.
Let $T' = T \bs \{n-1\}$. 
Then $(\{i\}, T') \der{c} (\{i\}, T)$ for $1 \le i \le m-2$.
Also, $(\{1\}, T) \der{a^2} (\{{m-1}\}, T)$ and $(\{2\}, T) \der{a^2} (\{0\}, T)$.
Therefore all states of the form $(\{i\}, T)$ with $|T| = k+1$, $0,n-1 \in T$ are reachable.
By the same argument as in Theorem \ref{thm:KRunionL}, all states of the form $(\{i\}, T)$ with $|T| = k+1$ are reachable.

Now suppose all states of the form $(S,T)$ are reachable for $|S| = k \ge 1$.
Again, it suffices to consider only the subsets $S \subseteq Q_1$ of size $k+1$ with $0, {m-1} \in S$,
and show these $(S,T)$ are reachable. 
Let $S' = S \bs \{{m-1}\}$; then $S' \der{c} S$.
If $0$ and $n-1$ are both in $T$ or both not in $T$, then $T \der{c} T$; hence $(S', T) \der{c} (S,T)$. 
For the other $T$, we divide the problem into two cases.

\noin
{\bf Case 1:} $m$ is odd. 
Let $w \in \{a,b\}^*$ be a permutation of states on $\cN_1$ and $\cN_2$.
We show how to construct another word $w' \in \{a,b\}^*$ which performs the same transformation as $w$ on $\cN_2$, 
but maps $S$ to itself in $\cN_1$.
To do this, we make three changes to $w$:
\be
\item[(i)] Add $a^{m-1}$ to the beginning of $w$.
\item[(ii)] Replace all instances of $a$ in $w$ by $a^m$.
\item[(iii)] Add $a^{m+1}$ to the end of $w$.
\ee
Call the resulting word $w'$. Because $m$ is odd and $a^2 : \one_{Q_2}$ on $\cN_2$, $w'$ is the same transformation as $w$ on $\cN_2$.
Consider applying $w'$ to $S$. Change (i) maps $S$ to some $S'$ with $0, 1 \in S'$. 
Since both $a^m$ and $b$ map $S'$ to itself,  the transformation caused by change (ii) maps $S'$ to itself.
Finally, change (iii) is the inverse of (i), mapping $S'$ back to $S$.

For any state $T \subseteq Q_2$ of size $k+1$, there is a word $w \in \{a,b\}^*$ which permutes $T$ to $T'$, for some $T'$ of size $k+1$ and $0, n-1 \in T'$.
Using the above construction, $(S,T)$ is reachable from $(S,T')$ by some permutation word.
Therefore all $2^{m+n}$ states are reachable for odd $m$.
Since the two NFA's are symmetric,  the same argument applies for reachability of all states
if $n$ is odd.

\noin
{\bf Case 2:} $m$ and $n$ are both even.
Suppose first that $1 \in S$, and that $T$ is of the form $T = \{0, t_1,\dots, t_l\}$, $0 < t_1 < \cdots < t_l$.
Let $j = n-1 - t_l$,  $T' = \{0, t_1 + j, \dots, t_{l-1} + j, n-1\}$, and $w = (ab)^j$.
Since  $ab$ is the cycle $(n-1,n-2,\dots, 1)$ of length $n-1$ in $\cN_2$,
 $T' \der w T$.

Define the words $tr_i = a^i ba^{m-i}$ that act as the transpositions $(i,i+1)$ in $\cN_1$. 
They act in $\cN_2$ as $b$ if $i$ is even, and $aba$ if $i$ is odd.
Using the $tr_i$, we show how to construct $w' \in \{a,b\}^*$ from $w$ so that $T' \der{w'} T$ and $S \der{w'} S$. 
We may assume that $j$ is even, as if $j$ is odd the same transformation can be caused by $w = (ba)^{j+n-1}$.
Since $0, 1, {m-1} \in S$, $w' = (tr_0 tr_{m-1})^{j/2}$ maps $T'$ to $T$ and $S$ to itself, and is the desired transformation.
It follows that all states of the form $(S,T)$, $0, 1, {m-1} \in S$, $|S| = k+1$, $0 \in T$ are reachable.
From states of this form, any $T$ can be reached by applying cyclic shifts $b^j$, which map $S$ to itself.

Now suppose $i \in S$, $1 < i < m-1$, and this $i$ minimal. 
If $i$ is even, let $w = tr_{i-1} (tr_{m-1})^{n-1}$. 
Then there exists $S'$ of size $k+1$ containing $0, {m-1},$ and ${i-1}$ such that
$S' \der w S$.
Moreover, $w$ acts as $(aba)^n : \one_{Q_2}$ on $\cN_2$, so $(S', T) \der w (S,T)$ for all $T \subseteq Q_2$.
If $i$ is odd, let $w = tr_{i-1} tr_{i-2} tr_{i-1} tr_{m-1}$, which acts as the transformation $({i-2}, i)(0,{m-1})$ on $\cN_1$ and $(ba)^4$ in $\cN_2$.
Since $n-1$ is odd, applying $w^{n-1}$ is the same transformation on $\cN_1$, while becoming the identity on $\cN_2$ (as $ba$ causes a cycle of length $n-1$).
Applying it to $(S', T)$ for some $S'$ containing $0, {i-2}, {m-1}$ results in $(S,T)$.
It follows by induction on $i$ that all states $S \cup T$ with $|S| = k+1$, and $\{0, m-1\} \subsetneq S$ are reachable.

Finally, suppose $S = \{0, {m-1}\}$. 
If $m \ge 6$, applying $a^2$ does not change $\cN_2$, but maps $S$ to  $S'=\{m-2,m-3\}$; thus $0, 1, {m-1} \notin S'$. 
Reachability for all states of the form $S' \cup T$ follows from the same argument as the case $0, 1, {m-1} \in S$.
Since $n$ is even, $(S', T) \der{a^{n-2}} (S, T)$, and  all of these states are reachable as well.
By symmetry, this argument applies when $n \ge 6$.

The only case remaining is $m = n = 4$. Computation shows that only 232 of the possible 256 states are reachable.

Next we examine the distinguishability of the reachable states.
Let $(S_1,T_1)$ and $(S_2, T_2)$ be two distinct states of $\cP$,
with $S_1 \neq S_2$.
We may apply a cyclic shift $b^k$ if necessary so that for each $i = 1,2$, either
(1) $T_i \in\{\emp, Q_2\}$, or (2) $\emp \subsetneq T_i \cap \{0,1, \dots, n-2\} \subsetneq \{0,1,\dots, n-2\}$. 
This is possible because $n \ge 3$.
Applying a cyclic shift $a^l$ if necessary, we may assume that $0 \in S_1 \bs S_2$.
As in Theorem \ref{thm:KRunionL}, we map $S_1$ to $Q_1$ and $S_2$ to $\emp$ by applying $(ca^{m-1})^{m-2}$. 

If the $T_i$ are $\emp$ or $Q_2$, this transformation leaves them unchanged. 
Otherwise, by the above condition, they are not mapped to either $\emp$ or $Q_2$.
Therefore we can map any pair of states of the form $(S_1, T_1)$ and $(S_2, T_2)$, $S_1 \neq S_2$ to $(Q_1, T_1')$, $(\emp, T_2')$ with $T_i' \in \{\emp, Q_2\} \iff T_i \in \{\emp, Q_2\}$ for $i = 1,2$.
A similar claim holds for the case $T_1 \neq T_2$ by switching the $a$'s and $b$'s.

We now consider each of the boolean operations separately.
\smallskip

\noin
{\bf Union:} The states $(Q_1, T)$ and $(S, Q_2)$ are final for all possible $S$ and $T$, 
and are all indistinguishable because any input leads to a state of the same form.

We now consider the $(2^m - 1)(2^n - 1)$ states not containing $Q_1$ or $Q_2$, and show they are all distinguishable.
By the above claim, and since the two DFA's are symmetric, we can reduce all pairs to the form 
$(Q_1, T_1)$, $(\emp, T_2)$, where $T_1, T_2 \neq Q_2$. 
These states are distinguishable by applying a cyclic shift $b^k$ mapping $T_2$ to a non-final state.
\smallskip

\noin
{\bf Intersection:} The states $(\emp, T)$ and $(S, \emp)$ are non-final and indistinguishable for all possible $S$ and $T$.
By the above claim again, all other states (not containing an $\emp$) can be reduced to the case $(Q_1, T_1)$, $(\emp, T_2)$, $T_1, T_2 \neq \emp$. 
Mapping $T_1$ to a final state using a cyclic shift will distinguish the states.
\smallskip

\noin
{\bf Difference:} We consider the operation $U_m^R \bs U_n^R$. 
The indistinguishable states are those of the form $(\emp, T)$ and $(S, Q_2)$, which are all non-final.
For $(Q_1, T_1)$, $(\emp, T_2)$ are distinguished by shifting $T_1$ to a non-final state, and $(S_1, Q_2)$, $(S_2, \emp)$ are distinguished by shifting $S_2$ to a final state.
\smallskip

\noin
{\bf Symmetric difference}
We first note that $(S, T)$ is final if and only if $(\bar S, \bar T)$ if final. 
Moreover, one can verify that if two states are complementary, then they are mapped to complementary states under any input.
Therefore $(S, T)$ and $(\bar S, \bar T)$ are indistinguishable.
This leads to a maximum of $2^{n+m-1}$ distinguishable states.

For any state $(S,T)$, either $S$ or $\bar S$ contains $q_0$. 
Therefore to complete the proof, we only need to show that all states of the form $(S,T)$ with $q_0 \in S$ are distinguishable. 
Let $(S_1, T_1)$ and $(S_2, T_2)$ be two such states.
If $T_1 = T_2$, then $S_1 \neq S_2$, there exists $q_k$ such that $k \in S_1 \oplus S_2$, and hence $a^k$ distinguishes the states.
If $T_1 \neq T_2$, by applying $b^2$ if necessary, we may assume that there exists $k \in \{0,\dots, n-2\}$ such that $k \in T_1 \oplus T_2$.
By applying $ca^{m-1}$, we may assume that $q_0, q_1 \in S_1 \cap S_2$.
This does not change the fact that $T_1$ and $T_2$ are distinct, by the above assumption.
So then applying $b^k$ for $k \in T_1 \oplus T_2$ distinguishes the two states.
\qed
\end{proof}

For $m\ge 3$, let  $\cU_{\{0,2\},m}(a,b,c)$  be the DFA obtained from $\cU_{m}(a,b,c)$ 
 by changing the set of final states  to $\{0,2\}$.
For $n\ge 4$, let $\cU_{\{1,3\},n}(b,a,c)$ $(\{1,3\})$ be the DFA obtained from $\cU_{n}(b,a,c)$
 by changing the set of final states  to $\{1,3\}$, and for $n=3$, use $\cU_{\{1\},n}(b,a,c)$ with final state 3.

\begin{theorem}[$K^R\circ L^R$]
Let $K_m=U_{\{0,2\},m}(a,b,c)$ and $L_n=U_{\{1,3\},n}(b,a,c)$ for $n\ge 4$ and
let $L_3=U_{\{1\},3\}}$.
For $m,n \ge 3$, the complexities of $K_m^R\cup L_n^R$, $K_m^R\cap L_n^R$, and
$K_m^R\setminus L_n^R$ 
are $(2^m-1)(2^n-1) + 1$, whereas that of 
$K_m^R\oplus L_n^R$ is
 $2^{m+n-1}$.
\end{theorem}
\begin{proof}
If it is not the case $m=n=4$, then by Lemma~\ref{lem:2rev}, it suffices to show that  state $(\{m-1\}, \{n-1\})$ is reachable from the initial state of the NFA.
If $n = 3$, the initial state is $(\{0,2\}, \{1\})$. We have the chain $(\{0,2\}, \{1\}) \der{ab^2c} (\{1\}, \{1\}) \der{a^2b^2} (\{m-1\}, \{n-1\})$.

Suppose $n \ge 4$. The initial state is $(\{0,2\}, \{1,3\})$. Apply the following: $(\{0,2\}, \{1,3\}) \der{ac} (\{1\}, \{0,3,n-1\}) \der{a^3} (\{m-2\}, \{1,3,n-1\})$.
If $n = 4$, then $n-1 = 3$, and we can apply $(\{m-2\}, \{1,3\}) \der c (\{m-2\}, \{1\}) \der{b^2a^{m-1}} (\{m-1\}, \{n-1\})$.
If $n > 4$, then apply $(\{m-2\}, \{1,3,n-1\}) \der{cb^2c} (\{m-2\}, \{1\}) \der{b^2a^{m-1}} (\{m-1\}, \{n-1\})$.

For every case except $m=n=4$, this shows that all states are reachable.
When $m=n=4$, one can verify through explicit enumeration that the states unreachable from $(\{3\}, \{3\})$ are exactly the states reached from $(\{0,2\}, \{1,3\})$ by words in $\{a,b\}^*$. 
Therefore in this case all states are reachable as well.
\qed
\end{proof}

\section{Product and Star}
\label{sec:prodstar}
\subsection{The Language $KL^R$}
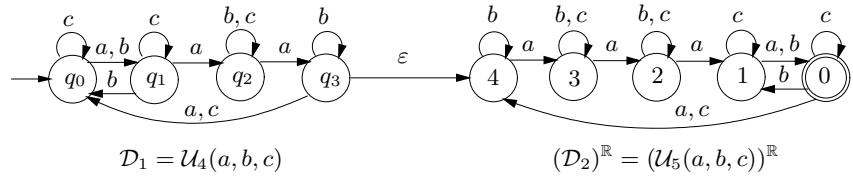
\begin{figure}[hbt]
\begin{center}
\setlength{\unitlength}{0.00039370in}
\begingroup\makeatletter\ifx\SetFigFont\undefined%
\gdef\SetFigFont#1#2#3#4#5{%
  \reset@font\fontsize{#1}{#2pt}%
  \fontfamily{#3}\fontseries{#4}\fontshape{#5}%
  \selectfont}%
\fi\endgroup%
{\renewcommand{\dashlinestretch}{30}
\begin{picture}(11082,2292)(0,-10)
\put(1437,109){\makebox(0,0)[lb]{\smash{{\SetFigFont{9}{10.8}{\rmdefault}{\mddefault}{\updefault}$\cD_1=\cU_4(a,b,c)$}}}}
\put(1894.500,1682.929){\arc{394.717}{2.4948}{6.9299}}
\blacken\thicklines
\path(2054.033,1704.097)(2052.000,1564.000)(2125.998,1682.977)(2054.033,1704.097)
\thinlines
\put(4181.500,1690.929){\arc{394.717}{2.4948}{6.9299}}
\blacken\thicklines
\path(4341.033,1712.097)(4339.000,1572.000)(4412.998,1690.977)(4341.033,1712.097)
\thinlines
\put(806.500,1675.929){\arc{394.717}{2.4948}{6.9299}}
\blacken\thicklines
\path(966.033,1697.097)(964.000,1557.000)(1037.998,1675.977)(966.033,1697.097)
\thinlines
\put(7429.500,1698.929){\arc{394.717}{2.4948}{6.9299}}
\blacken\thicklines
\path(7589.033,1720.097)(7587.000,1580.000)(7660.998,1698.977)(7589.033,1720.097)
\thinlines
\put(6380.500,1683.929){\arc{394.717}{2.4948}{6.9299}}
\blacken\thicklines
\path(6540.033,1705.097)(6538.000,1565.000)(6611.998,1683.977)(6540.033,1705.097)
\thinlines
\put(8524.500,1690.929){\arc{394.717}{2.4948}{6.9299}}
\blacken\thicklines
\path(8684.033,1712.097)(8682.000,1572.000)(8755.998,1690.977)(8684.033,1712.097)
\thinlines
\put(9642.500,1705.929){\arc{394.717}{2.4948}{6.9299}}
\blacken\thicklines
\path(9802.033,1727.097)(9800.000,1587.000)(9873.998,1705.977)(9802.033,1727.097)
\thinlines
\put(10767.500,1690.929){\arc{394.717}{2.4948}{6.9299}}
\blacken\thicklines
\path(10927.033,1712.097)(10925.000,1572.000)(10998.998,1690.977)(10927.033,1712.097)
\thinlines
\put(817,1255){\ellipse{630}{630}}
\put(4164,1274){\ellipse{630}{630}}
\put(3035,1286){\ellipse{630}{630}}
\put(1900,1250){\ellipse{630}{630}}
\put(8522,1280){\ellipse{630}{630}}
\put(9643,1287){\ellipse{630}{630}}
\put(7429,1290){\ellipse{630}{630}}
\put(6372,1272){\ellipse{630}{630}}
\put(10758,1277){\ellipse{540}{540}}
\put(10759,1277){\ellipse{630}{630}}
\path(2142,1474)(2772,1474)
\blacken\thicklines
\path(2637.000,1436.500)(2772.000,1474.000)(2637.000,1511.500)(2637.000,1436.500)
\thinlines
\path(3298,1489)(3928,1489)
\blacken\thicklines
\path(3793.000,1451.500)(3928.000,1489.000)(3793.000,1526.500)(3793.000,1451.500)
\thinlines
\path(1017,1481)(1647,1481)
\blacken\thicklines
\path(1512.000,1443.500)(1647.000,1481.000)(1512.000,1518.500)(1512.000,1443.500)
\thinlines
\path(1655,1069)(1070,1069)
\blacken\thicklines
\path(1205.000,1106.500)(1070.000,1069.000)(1205.000,1031.500)(1205.000,1106.500)
\thinlines
\path(12,1264)(499,1264)
\blacken\thicklines
\path(379.000,1234.000)(499.000,1264.000)(379.000,1294.000)(379.000,1234.000)
\thinlines
\path(7691,1505)(8276,1505)
\blacken\thicklines
\path(8141.000,1467.500)(8276.000,1505.000)(8141.000,1542.500)(8141.000,1467.500)
\thinlines
\path(8787,1497)(9372,1497)
\blacken\thicklines
\path(9237.000,1459.500)(9372.000,1497.000)(9237.000,1534.500)(9237.000,1459.500)
\thinlines
\path(6597,1512)(7182,1512)
\blacken\thicklines
\path(7047.000,1474.500)(7182.000,1512.000)(7047.000,1549.500)(7047.000,1474.500)
\thinlines
\path(9912,1497)(10497,1497)
\blacken\thicklines
\path(10362.000,1459.500)(10497.000,1497.000)(10362.000,1534.500)(10362.000,1459.500)
\thinlines
\path(4497,1279)(6027,1279)
\blacken\thicklines
\path(5892.000,1241.500)(6027.000,1279.000)(5892.000,1316.500)(5892.000,1241.500)
\thinlines
\path(10521,1129)(9936,1129)
\blacken\thicklines
\path(10071.000,1166.500)(9936.000,1129.000)(10071.000,1091.500)(10071.000,1166.500)
\thinlines
\path(3919,1039)(3918,1039)(3916,1038)
	(3913,1036)(3907,1034)(3899,1030)
	(3888,1025)(3874,1019)(3857,1012)
	(3837,1003)(3814,993)(3788,982)
	(3759,970)(3727,957)(3693,943)
	(3657,929)(3619,914)(3579,899)
	(3537,884)(3494,868)(3449,853)
	(3403,838)(3356,823)(3308,808)
	(3258,794)(3207,780)(3154,767)
	(3099,754)(3042,742)(2983,731)
	(2922,721)(2858,711)(2792,702)
	(2723,695)(2652,689)(2578,684)
	(2503,681)(2427,679)(2347,679)
	(2269,682)(2193,686)(2120,692)
	(2050,699)(1983,707)(1920,717)
	(1859,727)(1801,739)(1746,751)
	(1693,764)(1642,777)(1593,791)
	(1545,806)(1499,821)(1455,836)
	(1412,851)(1371,867)(1331,883)
	(1294,898)(1258,913)(1224,927)
	(1192,941)(1164,954)(1137,966)
	(1114,977)(1094,986)(1077,995)
	(1063,1001)(1052,1007)(1043,1011)(1031,1017)
\blacken\thicklines
\path(1168.518,990.167)(1031.000,1017.000)(1134.977,923.085)(1168.518,990.167)
\thinlines
\path(10625,995)(10624,995)(10623,994)
	(10620,993)(10615,992)(10609,989)
	(10600,986)(10588,982)(10574,977)
	(10557,972)(10537,965)(10515,957)
	(10489,949)(10460,939)(10429,929)
	(10395,918)(10358,906)(10320,894)
	(10279,881)(10236,869)(10191,855)
	(10145,842)(10097,828)(10047,815)
	(9997,801)(9945,788)(9892,774)
	(9837,761)(9781,748)(9724,736)
	(9665,723)(9605,712)(9543,700)
	(9480,689)(9414,679)(9346,669)
	(9276,659)(9204,651)(9129,642)
	(9051,635)(8971,629)(8889,623)
	(8804,619)(8717,615)(8629,613)
	(8540,612)(8447,613)(8356,615)
	(8266,618)(8178,623)(8094,628)
	(8012,635)(7933,643)(7858,651)
	(7785,660)(7714,670)(7647,680)
	(7581,691)(7518,703)(7457,715)
	(7398,727)(7340,740)(7284,753)
	(7229,767)(7176,780)(7125,794)
	(7074,808)(7025,822)(6978,836)
	(6933,850)(6889,864)(6847,877)
	(6807,890)(6770,902)(6735,914)
	(6702,925)(6673,936)(6646,945)
	(6622,953)(6601,961)(6583,967)
	(6569,973)(6557,977)(6547,981)
	(6540,983)(6530,987)
\blacken\thicklines
\path(6669.272,971.680)(6530.000,987.000)(6641.417,902.044)(6669.272,971.680)
\put(716,1197){\makebox(0,0)[lb]{\smash{{\SetFigFont{9}{10.8}{\rmdefault}{\mddefault}{\updefault}$q_0$}}}}
\put(1790,1211){\makebox(0,0)[lb]{\smash{{\SetFigFont{9}{10.8}{\rmdefault}{\mddefault}{\updefault}$q_1$}}}}
\put(2922,1225){\makebox(0,0)[lb]{\smash{{\SetFigFont{9}{10.8}{\rmdefault}{\mddefault}{\updefault}$q_2$}}}}
\put(1826,1975){\makebox(0,0)[lb]{\smash{{\SetFigFont{9}{10.8}{\familydefault}{\mddefault}{\updefault}$c$}}}}
\put(2292,769){\makebox(0,0)[lb]{\smash{{\SetFigFont{9}{10.8}{\familydefault}{\mddefault}{\updefault}$a,c$}}}}
\put(3538,1601){\makebox(0,0)[lb]{\smash{{\SetFigFont{9}{10.8}{\familydefault}{\mddefault}{\updefault}$a$}}}}
\put(1271,1151){\makebox(0,0)[lb]{\smash{{\SetFigFont{9}{10.8}{\familydefault}{\mddefault}{\updefault}$b$}}}}
\put(688,1961){\makebox(0,0)[lb]{\smash{{\SetFigFont{9}{10.8}{\familydefault}{\mddefault}{\updefault}$c$}}}}
\put(2811,1998){\makebox(0,0)[lb]{\smash{{\SetFigFont{9}{10.8}{\familydefault}{\mddefault}{\updefault}$b,c$}}}}
\put(4084,1202){\makebox(0,0)[lb]{\smash{{\SetFigFont{9}{10.8}{\rmdefault}{\mddefault}{\updefault}$q_3$}}}}
\put(1116,1593){\makebox(0,0)[lb]{\smash{{\SetFigFont{9}{10.8}{\familydefault}{\mddefault}{\updefault}$a,b$}}}}
\put(2367,1594){\makebox(0,0)[lb]{\smash{{\SetFigFont{9}{10.8}{\familydefault}{\mddefault}{\updefault}$a$}}}}
\put(8456,1200){\makebox(0,0)[lb]{\smash{{\SetFigFont{9}{10.8}{\rmdefault}{\mddefault}{\updefault}$2$}}}}
\put(7850,1634){\makebox(0,0)[lb]{\smash{{\SetFigFont{9}{10.8}{\familydefault}{\mddefault}{\updefault}$a$}}}}
\put(8973,1633){\makebox(0,0)[lb]{\smash{{\SetFigFont{9}{10.8}{\familydefault}{\mddefault}{\updefault}$a$}}}}
\put(7361,1170){\makebox(0,0)[lb]{\smash{{\SetFigFont{9}{10.8}{\rmdefault}{\mddefault}{\updefault}$3$}}}}
\put(6288,1177){\makebox(0,0)[lb]{\smash{{\SetFigFont{9}{10.8}{\rmdefault}{\mddefault}{\updefault}$4$}}}}
\put(9582,1207){\makebox(0,0)[lb]{\smash{{\SetFigFont{9}{10.8}{\rmdefault}{\mddefault}{\updefault}$1$}}}}
\put(10661,1207){\makebox(0,0)[lb]{\smash{{\SetFigFont{9}{10.8}{\rmdefault}{\mddefault}{\updefault}$0$}}}}
\put(7183,2015){\makebox(0,0)[lb]{\smash{{\SetFigFont{9}{10.8}{\familydefault}{\mddefault}{\updefault}$b,c$}}}}
\put(6747,1626){\makebox(0,0)[lb]{\smash{{\SetFigFont{9}{10.8}{\familydefault}{\mddefault}{\updefault}$a$}}}}
\put(4056,1983){\makebox(0,0)[lb]{\smash{{\SetFigFont{9}{10.8}{\familydefault}{\mddefault}{\updefault}$b$}}}}
\put(8278,2022){\makebox(0,0)[lb]{\smash{{\SetFigFont{9}{10.8}{\familydefault}{\mddefault}{\updefault}$b,c$}}}}
\put(9957,1623){\makebox(0,0)[lb]{\smash{{\SetFigFont{9}{10.8}{\familydefault}{\mddefault}{\updefault}$a,b$}}}}
\put(10133,1220){\makebox(0,0)[lb]{\smash{{\SetFigFont{9}{10.8}{\familydefault}{\mddefault}{\updefault}$b$}}}}
\put(8765,769){\makebox(0,0)[lb]{\smash{{\SetFigFont{9}{10.8}{\familydefault}{\mddefault}{\updefault}$a,c$}}}}
\put(9522,2015){\makebox(0,0)[lb]{\smash{{\SetFigFont{9}{10.8}{\familydefault}{\mddefault}{\updefault}$c$}}}}
\put(10683,2009){\makebox(0,0)[lb]{\smash{{\SetFigFont{9}{10.8}{\familydefault}{\mddefault}{\updefault}$c$}}}}
\put(6289,2010){\makebox(0,0)[lb]{\smash{{\SetFigFont{9}{10.8}{\familydefault}{\mddefault}{\updefault}$b$}}}}
\put(5104,1474){\makebox(0,0)[lb]{\smash{{\SetFigFont{9}{10.8}{\familydefault}{\mddefault}{\updefault}$\eps$}}}}
\put(7144,109){\makebox(0,0)[lb]{\smash{{\SetFigFont{9}{10.8}{\rmdefault}{\mddefault}{\updefault}$(\cD_2)^\rev=(\cU_5(a,b,c))^\rev$}}}}
\thinlines
\put(3041.500,1712.929){\arc{394.717}{2.4948}{6.9299}}
\blacken\thicklines
\path(3201.033,1734.097)(3199.000,1594.000)(3272.998,1712.977)(3201.033,1734.097)
\end{picture}
}
\end{center}
\caption{NFA $\cN$ for $U_4(a,b,c)(U_5(a,b,c))^R$.} 
\label{fig:KLr}
\end{figure}

Cui, Gao, Kari and Yu showed in~\cite{CGKY_IJFCS12} that the complexity of $KL^R$ is  
$(m-1)2^n+2^{n-1}-(m-1)$,   with ternary witnesses.
We now prove that the bound can also be met by one stream.
The NFA $\cN$ for $U_4(a,b,c)(U_5(a,b,c))^R$ is shown in Fig.~\ref{fig:KLr}.

\begin{theorem}
\label{thm:KLR}
For $m,n \ge 3$,  the  complexity of  the product $U_m(a,b,c)(U_n(a,b,c))^R$ is $(m-1)2^n+2^{n-1}-(m-1)$.
\end{theorem}
\begin{proof}
Let $\cD_1 = (Q_1, \Sigma, \delta_1, q_0, \{q_{m-1}\})$ and $\cD_2 = (Q_2, \Sigma, \delta_2, 0, \{n-1\})$ be the minimal DFA's of $U_m(a,b,c)$  and $U_n(a,b,c)$, where 
$Q_1=\{q_0,\ldots,q_{m-1}\}$ and $Q_2=\{0,\ldots,n-1\}$.
Let $\cN_2$ be $\cD_2^\rev$, and let $\cN$ be the NFA for the product of $\cD_1$ and $\cN_2$, as illustrated in Fig.~\ref{fig:KLr}.

We use the subset construction on $\cN$ to get a DFA $\cP$ for this product.
Any state of $\cP$ must either not contain $q_{m-1}$, or contain both $q_{m-1}$ and $n-1$. 
There are $(m-1)2^n$ states of the former type, and $2^{n-1}$ states of the latter. 
We will show that all of these states are reachable.

Set $\{q_0\}$ is initial,  $\{q_i\}$ is reached by $a^i$, for $i=1,\ldots,m-2$, and
$\{q_{m-1},n-1\}$  by $a^{m-1}$.
Also, $\{q_{m-1},n-1\} \der a \{q_0,n-2\}$, and from there
$\{q_0,j\}$ is reached by $(ab)^{n-2-j}$
for $j = 2,\ldots,n-3$, $\{q_0,0\}$ by $(ab)^{n-3}$, and $\{q_0,n-1\}$ by $(ab)^{n-2}$.

If $n \nmid m$, then $\{1 + m \pmod n\} \neq \{1\}$ and $\{q_0, 1\}$ is reachable by $a^m$ from $\{q_0,1+m \pmod n\}$. 
If $n \mid m$, then $m-1 \equiv n-1 \pmod n$; so applying $a^{m-1}c$ sends $\{q_0, 0\}$ to $\{q_0, 1\}$.
For $i=1,\ldots,m-2$, $\{q_i,j\}$ is reached from $\{q_{0}, i+j \pmod n\}$ by $a^i$.
So all states  $\{q_i\}\cup S$, where $i<m-1$ and $|S| \le 1$ are reachable.

For the rest of the proof $S$ and $T$ will denote subsets of $Q_2$.
Suppose it is possible to reach all states of the form $\{q_i\} \cup S$, where
$i<m-1$, $S\subseteq Q_2$, and $|S| = k \ge 1$. 
We want to show it is possible to reach all states of the form $\{q_{m-1}\} \cup T$, $|T| = k+1$, and $n-1 \in T$. 
Let $T = \{t_1, \dots, t_k, n-1\}$.
Then $\{q_{m-2},(t_1+1), \ldots, (t_k+1)\} \der a \{q_{m-1}\} \cup T$, and this state is reachable.

Now suppose all  states of the form $\{q_{m-1}\} \cup T$, where $|T| = k \ge 2$ and $n-1 \in T$ are reachable. 
We want to show that all states of the form $\{q_i\} \cup S$ with $|S|=k$ are reachable. 
Applying $a$ shows that all states of the form $\{q_0\} \cup T$ with $|T| = k$ and $n-2 \in T$ are reachable.
The word $ab$ sends $q_0$ to $q_0$, and acts as the cycle $(n-1, n-2, \dots, 2, 0)$ on the states of $\cN_2$. 
Hence for any subset $T' \subseteq Q_2$ of size $k \ge 2$, there exists an integer $d$ and $T \subseteq Q_2$ containing $n-2$ such that $T \der{(ab)^d} T'$.
Therefore all states of the form $\{q_0\} \cup S$ with $|S| = k$ are reachable.
Let $i < m-1$ and $S = \{s_1, \dots, s_k\} \subseteq Q_2$. 
State $\{q_i\} \cup S$ is reachable by $a^i$ from  state $\{q_0\} \cup \{s_1 + i, \dots, s_k + i\}$,
where addition is modulo $n$. Hence all states of the form $\{q_i\} \cup S$ with $i < m-1$ and $|S| = k+1$ are reachable.

Combining these two results shows that all the required states are reachable.

\vspace{5pt}

For distinguishability, first note that all $m$ states of the form $\{q_i\} \cup Q_2$ are final and indistinguishable.

Suppose we have two states $\{q_i\} \cup S$ and $\{q_j\} \cup T$ with $S \neq T$. 
Let $k \in S \oplus T$; then $a^k$ distinguishes the two states.
Now consider the pair $\{q_i\} \cup S$, $\{q_j\} \cup S$, $S \neq Q_2$.
Let $k \notin S$, and apply $a^k$ to get $\{q_{i'}\} \cup S'$, $\{q_{j'}\} \cup T'$.
If $S' \neq T'$, then by the previous argument the states are distinguishable.
Otherwise, $S' = T'$ and $0 \notin S'$.
So without loss of generality we may assume that $0 \notin S$.
We know that $ba$ acts as the cycle $(q_0, q_2, q_3, \dots, q_{m-1})$ on $\cD_1$, and maps only 0 to 0 in $\cN_2$.
Since $i \neq j$, at least one of $i,j$ is not equal to 1.
Then by applying some $(ba)^d$ if necessary, we may assume that $i < m-2$, $j = m-2$. 
Apply $a$ to get $\{q_{i+1}\} \cup T$, $\{q_{m-1}\} \cup T \cup \{n-1\}$, where $n-1 \notin T$.
Since these states contain different subsets of $Q_2$, they are distinguishable by the previous argument.
\qed
\end{proof}

\subsection{The Language $K^RL$}

Let $\cV_n(a,b,c,d)=(Q_\cV,\Sig,\delta_\cV,0, \{n-1\})$, where $Q=\{0,\ldots,n-1\}$,
$a:(0,\ldots,n-1)$,  $b:(n-2,n-1)$,
$c:{n-1\choose n-2}$, and
$d:\one_{Q_n}$.
Let $V_n(a,b,c,d)$ be the language of $\cV_n(a,b,c,d)$.

It was shown in~\cite{CGKY_TCS12} by Cui, Gao, Kari and Yu that
$3\cdot2^{m+n-2}$ is a tight bound 
for $K_m^RL_n$. 
They used $\cV_n(a,b,c,d)$ as witnesses $K_m$  (some relabelling is needed),
and  $L_n$
with $a,c:\one_{Q_n}$, $b:{Q_n\choose 0}$,   $d:(0,\ldots,n-1)$ and final state $n-1$. 
We prove that the permutationally equivalent dialects $(V_m(a,b,c,d)\mid m\ge 3)$ and $(V_n(d,c,b,a)\mid n\ge 3)$ can also be used.

\begin{figure}[hbt]
\begin{center}
\setlength{\unitlength}{0.00039370in}
\begingroup\makeatletter\ifx\SetFigFont\undefined%
\gdef\SetFigFont#1#2#3#4#5{%
  \reset@font\fontsize{#1}{#2pt}%
  \fontfamily{#3}\fontseries{#4}\fontshape{#5}%
  \selectfont}%
\fi\endgroup%
{\renewcommand{\dashlinestretch}{30}
\begin{picture}(11084,2323)(0,-10)
\put(7452,124){\makebox(0,0)[lb]{\smash{{\SetFigFont{9}{10.8}{\rmdefault}{\mddefault}{\updefault}$\cD_2=\cV_5(d,c,b,a)$}}}}
\put(1894.500,1705.929){\arc{394.717}{2.4948}{6.9299}}
\blacken\thicklines
\path(2054.033,1727.097)(2052.000,1587.000)(2125.998,1705.977)(2054.033,1727.097)
\thinlines
\put(4181.500,1713.929){\arc{394.717}{2.4948}{6.9299}}
\blacken\thicklines
\path(4341.033,1735.097)(4339.000,1595.000)(4412.998,1713.977)(4341.033,1735.097)
\thinlines
\put(806.500,1698.929){\arc{394.717}{2.4948}{6.9299}}
\blacken\thicklines
\path(966.033,1720.097)(964.000,1580.000)(1037.998,1698.977)(966.033,1720.097)
\thinlines
\put(7429.500,1721.929){\arc{394.717}{2.4948}{6.9299}}
\blacken\thicklines
\path(7589.033,1743.097)(7587.000,1603.000)(7660.998,1721.977)(7589.033,1743.097)
\thinlines
\put(6380.500,1706.929){\arc{394.717}{2.4948}{6.9299}}
\blacken\thicklines
\path(6540.033,1728.097)(6538.000,1588.000)(6611.998,1706.977)(6540.033,1728.097)
\thinlines
\put(8524.500,1713.929){\arc{394.717}{2.4948}{6.9299}}
\blacken\thicklines
\path(8684.033,1735.097)(8682.000,1595.000)(8755.998,1713.977)(8684.033,1735.097)
\thinlines
\put(9642.500,1728.929){\arc{394.717}{2.4948}{6.9299}}
\blacken\thicklines
\path(9802.033,1750.097)(9800.000,1610.000)(9873.998,1728.977)(9802.033,1750.097)
\thinlines
\put(10767.500,1713.929){\arc{394.717}{2.4948}{6.9299}}
\blacken\thicklines
\path(10927.033,1735.097)(10925.000,1595.000)(10998.998,1713.977)(10927.033,1735.097)
\thinlines
\put(817,1278){\ellipse{630}{630}}
\put(4164,1297){\ellipse{630}{630}}
\put(3035,1309){\ellipse{630}{630}}
\put(8522,1303){\ellipse{630}{630}}
\put(9643,1310){\ellipse{630}{630}}
\put(7429,1313){\ellipse{630}{630}}
\put(10758,1300){\ellipse{540}{540}}
\put(10761,1299){\ellipse{630}{630}}
\put(6372,1295){\ellipse{630}{630}}
\put(1900,1273){\ellipse{630}{630}}
\path(12,1287)(499,1287)
\blacken\thicklines
\path(379.000,1257.000)(499.000,1287.000)(379.000,1317.000)(379.000,1257.000)
\thinlines
\path(4504,1302)(6012,1302)
\blacken\thicklines
\path(5877.000,1264.500)(6012.000,1302.000)(5877.000,1339.500)(5877.000,1264.500)
\blacken\path(2608.000,1444.500)(2743.000,1482.000)(2608.000,1519.500)(2608.000,1444.500)
\thinlines
\path(2743,1482)(2158,1482)
\blacken\thicklines
\path(3763.000,1452.500)(3898.000,1490.000)(3763.000,1527.500)(3763.000,1452.500)
\thinlines
\path(3898,1490)(3313,1490)
\blacken\thicklines
\path(1520.000,1452.500)(1655.000,1490.000)(1520.000,1527.500)(1520.000,1452.500)
\thinlines
\path(1655,1490)(1070,1490)
\blacken\thicklines
\path(7048.000,1482.500)(7183.000,1520.000)(7048.000,1557.500)(7048.000,1482.500)
\thinlines
\path(7183,1520)(6598,1520)
\blacken\thicklines
\path(8143.000,1482.500)(8278.000,1520.000)(8143.000,1557.500)(8143.000,1482.500)
\thinlines
\path(8278,1520)(7693,1520)
\blacken\thicklines
\path(9231.000,1474.500)(9366.000,1512.000)(9231.000,1549.500)(9231.000,1474.500)
\thinlines
\path(9366,1512)(8781,1512)
\blacken\thicklines
\path(10356.000,1482.500)(10491.000,1520.000)(10356.000,1557.500)(10356.000,1482.500)
\thinlines
\path(10491,1520)(9906,1520)
\blacken\thicklines
\path(10047.000,1159.500)(9912.000,1122.000)(10047.000,1084.500)(10047.000,1159.500)
\thinlines
\path(9912,1122)(10497,1122)
\path(1640,1100)(1092,1100)
\blacken\thicklines
\path(1227.000,1137.500)(1092.000,1100.000)(1227.000,1062.500)(1227.000,1137.500)
\thinlines
\path(3919,1062)(3918,1062)(3916,1061)
	(3913,1059)(3907,1057)(3899,1053)
	(3888,1048)(3874,1042)(3857,1035)
	(3837,1026)(3814,1016)(3788,1005)
	(3759,993)(3727,980)(3693,966)
	(3657,952)(3619,937)(3579,922)
	(3537,907)(3494,891)(3449,876)
	(3403,861)(3356,846)(3308,831)
	(3258,817)(3207,803)(3154,790)
	(3099,777)(3042,765)(2983,754)
	(2922,744)(2858,734)(2792,725)
	(2723,718)(2652,712)(2578,707)
	(2503,704)(2427,702)(2347,702)
	(2269,705)(2193,709)(2120,715)
	(2050,722)(1983,730)(1920,740)
	(1859,750)(1801,762)(1746,774)
	(1693,787)(1642,800)(1593,814)
	(1545,829)(1499,844)(1455,859)
	(1412,874)(1371,890)(1331,906)
	(1294,921)(1258,936)(1224,950)
	(1192,964)(1164,977)(1137,989)
	(1114,1000)(1094,1009)(1077,1018)
	(1063,1024)(1052,1030)(1043,1034)(1031,1040)
\blacken\thicklines
\path(1168.518,1013.167)(1031.000,1040.000)(1134.977,946.085)(1168.518,1013.167)
\thinlines
\path(10625,1018)(10624,1018)(10623,1017)
	(10620,1016)(10615,1015)(10609,1012)
	(10600,1009)(10588,1005)(10574,1000)
	(10557,995)(10537,988)(10515,980)
	(10489,972)(10460,962)(10429,952)
	(10395,941)(10358,929)(10320,917)
	(10279,904)(10236,892)(10191,878)
	(10145,865)(10097,851)(10047,838)
	(9997,824)(9945,811)(9892,797)
	(9837,784)(9781,771)(9724,759)
	(9665,746)(9605,735)(9543,723)
	(9480,712)(9414,702)(9346,692)
	(9276,682)(9204,674)(9129,665)
	(9051,658)(8971,652)(8889,646)
	(8804,642)(8717,638)(8629,636)
	(8540,635)(8447,636)(8356,638)
	(8266,641)(8178,646)(8094,651)
	(8012,658)(7933,666)(7858,674)
	(7785,683)(7714,693)(7647,703)
	(7581,714)(7518,726)(7457,738)
	(7398,750)(7340,763)(7284,776)
	(7229,790)(7176,803)(7125,817)
	(7074,831)(7025,845)(6978,859)
	(6933,873)(6889,887)(6847,900)
	(6807,913)(6770,925)(6735,937)
	(6702,948)(6673,959)(6646,968)
	(6622,976)(6601,984)(6583,990)
	(6569,996)(6557,1000)(6547,1004)
	(6540,1006)(6530,1010)
\blacken\thicklines
\path(6669.272,994.680)(6530.000,1010.000)(6641.417,925.044)(6669.272,994.680)
\put(1790,1234){\makebox(0,0)[lb]{\smash{{\SetFigFont{9}{10.8}{\rmdefault}{\mddefault}{\updefault}$q_2$}}}}
\put(4084,1225){\makebox(0,0)[lb]{\smash{{\SetFigFont{9}{10.8}{\rmdefault}{\mddefault}{\updefault}$q_0$}}}}
\put(6289,1223){\makebox(0,0)[lb]{\smash{{\SetFigFont{9}{10.8}{\rmdefault}{\mddefault}{\updefault}$0$}}}}
\put(7347,1230){\makebox(0,0)[lb]{\smash{{\SetFigFont{9}{10.8}{\rmdefault}{\mddefault}{\updefault}$1$}}}}
\put(9559,1231){\makebox(0,0)[lb]{\smash{{\SetFigFont{9}{10.8}{\rmdefault}{\mddefault}{\updefault}$3$}}}}
\put(10676,1231){\makebox(0,0)[lb]{\smash{{\SetFigFont{9}{10.8}{\rmdefault}{\mddefault}{\updefault}$4$}}}}
\put(7850,1657){\makebox(0,0)[lb]{\smash{{\SetFigFont{9}{10.8}{\familydefault}{\mddefault}{\updefault}$d$}}}}
\put(8973,1656){\makebox(0,0)[lb]{\smash{{\SetFigFont{9}{10.8}{\familydefault}{\mddefault}{\updefault}$d$}}}}
\put(5127,1423){\makebox(0,0)[lb]{\smash{{\SetFigFont{9}{10.8}{\familydefault}{\mddefault}{\updefault}$\eps$}}}}
\put(2381,1610){\makebox(0,0)[lb]{\smash{{\SetFigFont{9}{10.8}{\familydefault}{\mddefault}{\updefault}$a$}}}}
\put(2922,1248){\makebox(0,0)[lb]{\smash{{\SetFigFont{9}{10.8}{\rmdefault}{\mddefault}{\updefault}$q_1$}}}}
\put(8456,1223){\makebox(0,0)[lb]{\smash{{\SetFigFont{9}{10.8}{\rmdefault}{\mddefault}{\updefault}$2$}}}}
\put(6761,1655){\makebox(0,0)[lb]{\smash{{\SetFigFont{9}{10.8}{\familydefault}{\mddefault}{\updefault}$d$}}}}
\put(7026,2053){\makebox(0,0)[lb]{\smash{{\SetFigFont{9}{10.8}{\familydefault}{\mddefault}{\updefault}$a,b,c$}}}}
\put(9973,1668){\makebox(0,0)[lb]{\smash{{\SetFigFont{9}{10.8}{\familydefault}{\mddefault}{\updefault}$c,d$}}}}
\put(8825,762){\makebox(0,0)[lb]{\smash{{\SetFigFont{9}{10.8}{\familydefault}{\mddefault}{\updefault}$d$}}}}
\put(10653,2040){\makebox(0,0)[lb]{\smash{{\SetFigFont{9}{10.8}{\familydefault}{\mddefault}{\updefault}$a$}}}}
\put(9408,2045){\makebox(0,0)[lb]{\smash{{\SetFigFont{9}{10.8}{\familydefault}{\mddefault}{\updefault}$a,b$}}}}
\put(8149,2052){\makebox(0,0)[lb]{\smash{{\SetFigFont{9}{10.8}{\familydefault}{\mddefault}{\updefault}$a,b,c$}}}}
\put(5975,2045){\makebox(0,0)[lb]{\smash{{\SetFigFont{9}{10.8}{\familydefault}{\mddefault}{\updefault}$a,b,c$}}}}
\put(10002,1216){\makebox(0,0)[lb]{\smash{{\SetFigFont{9}{10.8}{\familydefault}{\mddefault}{\updefault}$b,c$}}}}
\put(732,2036){\makebox(0,0)[lb]{\smash{{\SetFigFont{9}{10.8}{\familydefault}{\mddefault}{\updefault}$d$}}}}
\put(1115,1616){\makebox(0,0)[lb]{\smash{{\SetFigFont{9}{10.8}{\familydefault}{\mddefault}{\updefault}$a,b$}}}}
\put(3537,1602){\makebox(0,0)[lb]{\smash{{\SetFigFont{9}{10.8}{\familydefault}{\mddefault}{\updefault}$a$}}}}
\put(2283,769){\makebox(0,0)[lb]{\smash{{\SetFigFont{9}{10.8}{\familydefault}{\mddefault}{\updefault}$a$}}}}
\put(1645,2028){\makebox(0,0)[lb]{\smash{{\SetFigFont{9}{10.8}{\familydefault}{\mddefault}{\updefault}$c,d$}}}}
\put(2647,2051){\makebox(0,0)[lb]{\smash{{\SetFigFont{9}{10.8}{\familydefault}{\mddefault}{\updefault}$b,c,d$}}}}
\put(3786,2052){\makebox(0,0)[lb]{\smash{{\SetFigFont{9}{10.8}{\familydefault}{\mddefault}{\updefault}$b,c,d$}}}}
\put(716,1220){\makebox(0,0)[lb]{\smash{{\SetFigFont{9}{10.8}{\rmdefault}{\mddefault}{\updefault}$q_3$}}}}
\put(1130,1175){\makebox(0,0)[lb]{\smash{{\SetFigFont{9}{10.8}{\familydefault}{\mddefault}{\updefault}$b,c$}}}}
\put(1265,109){\makebox(0,0)[lb]{\smash{{\SetFigFont{9}{10.8}{\rmdefault}{\mddefault}{\updefault}$(\cD_1)^\rev=(\cV_4(a,b,c,d))^\rev$}}}}
\thinlines
\put(3041.500,1735.929){\arc{394.717}{2.4948}{6.9299}}
\blacken\thicklines
\path(3201.033,1757.097)(3199.000,1617.000)(3272.998,1735.977)(3201.033,1757.097)
\end{picture}
}
\end{center}
\caption{NFA for $(V_4(a,b,c,d))^R\;V_5(d,c,b,a)$.} 
\label{fig:KrL}
\end{figure}
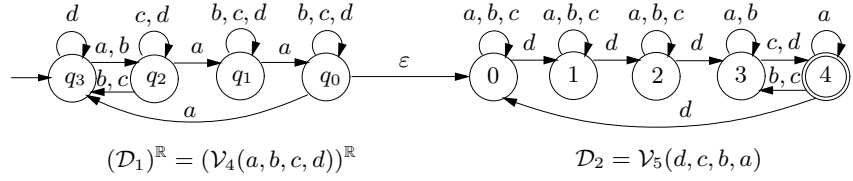

\begin{theorem}[$K_m^RL_n$]\mbox{}\\
\noin
For $m,n \ge 3$, the complexity of $(V_m(a,b,c,d))^R V_n(d,c,b,a)$ is $3\cdot 2^{m+n-2}$.
\end{theorem}
\begin{proof}
Let $\cD_1 = (Q_1, \Sigma, \delta_1, q_0, \{q_{m-1}\})$ and $\cD_2 = (Q_2, \Sigma, \delta_2, 0, \{n-1\})$ be the minimal DFA's of $V_m(a,b,c,d)$  and $V_n(d,c,b,a)$, where 
$Q_1=\{q_0,\ldots,q_{m-1}\}$ and $Q_2=\{0,\ldots,n-1\}$.
Let $\cN_1$ be $\cD_1^\rev$, and let $\cN$ be the NFA for the product of $\cN_1$ and $\cD_2$, as illustrated in Fig.~\ref{fig:KrL}.

We use the subset construction to get a DFA $\cP$ for this product.
We claim that all $2^{m+n-1}$  states of $\cP$  not containing $q_0$ and all $2^{m+n-2}$ states containing $q_0$ and $0$ are reachable.

The initial state is $\{q_{m-1}\}$. Then we have  $\{q_{m-1}\} \der{a^{m-1-i}} \{q_i\}$ for $i \ge 1$, and $\{q_{m-1}\} \der{a^{m-1}} \{q_0,0\}$.
Now suppose all states of the form $S \subsetneq Q_1 $, $|S| = k \ge 1$ are reachable.
Let $S = \{q_{s_1},\dots, q_{s_{k+1}}\}$ with $0 < s_1 < \cdots < s_{k+1}$. 
Let $i = s_{k+1} - s_k - 1$, and $j = m-1 - s_{k+1}$. 
Let $S' = \{q_{s_1+i+j},\dots, q_{s_{k-1} + i+j}, q_{m-2}\}$. Note that $S'$ is reachable.
Then $S$ is reachable by the sequence 
$$S' \der{c} S' \cup \{q_{m-1}\} \der{(ab)^i} \{q_{s_1 + j}, \dots, q_{s_{k-1} + j}, q_{s_k + j}, q_{m-1}\} \der{a^j} S.$$
On the other hand, setting $s_1 = 0$ shows the reachability for all states of the form $S \cup \{0\}$, $|S| = k+1$, $q_0 \in S$.

Suppose states of the form $S \cup T$ with $\emp \subsetneq S \subseteq Q_1 \bs \{q_0\}$, $T \subseteq Q_2$, and $|T| = k \ge 0$ are reachable.
Since $S$ is non-empty, $S \der{a^m} S \cup \{0\}$.
Let $T = \{t_1,\dots, t_{k+1}\}$, $t_1 < \cdots < t_{k+1}$.
Let $T' = \{t_2 - t_1,\dots, t_{k+1} - t_1\}$. 
By induction, $S \cup T'$ is reachable.
Then $S \cup T$ is reachable by the sequence $$S \cup T' \der{a^m} S \cup \{0\} \cup T' \der{d^{t_1}} S \cup T.$$
Moreover, if we take $S = \{q_{m-1}\}$, then $S \cup T \der{c^2} T$.

Finally, consider states of the form $S \cup T$ where $q_0 \in S$, $0 \in T$.
If $S \neq Q_1$, there exists an $S'$ with $q_0 \notin S'$ such that $S' \cup T \der{a^j} S \cup T$ for some $j$.
Note that $S' \cup T$ is reachable by the previous case.
If $S = Q_1$, then define $S' = Q_1 \bs \{q_0\}$.
Once again, $S' \cup T$ is reachable, and we have $S' \cup T \der{ac^2} S \cup T$.
Therefore all of the desired states are reachable.

We now prove that all of these states are distinguishable. 
Let $S_1 \cup T_1$, $S_2 \cup T_2$ be a pair of states, $S_1, S_2 \subseteq Q_1$, $T_1, T_2 \subseteq Q_2$.
If $T_1 \neq T_2$, then let $k \in T_1 \oplus T_2$. The states are distinguishable by $d^{n-1-k}$.
If $S_1 \neq S_2$ without loss of generality (applying a cyclic shift if necessary), assume $q_0 \in S_1 \oplus S_2$.
Applying $b^2$ ensures that $n-1 \notin T_1 \cup T_2$. 
Then applying $d$ transforms the pair to $S_1 \cup T_1'$, $S_2 \cup T_2'$, and $0 \in T_i'$ if and only if $q_0 \in S_i$.
So $T_1' \neq T_2'$, and the states are distinguishable.
\qed
\end{proof}

\subsection{The Language $(KL)^R=L^RK^R$}

Let $\cU_n(a,b,c,d)=(Q,\Sig,\delta_\cU,0, \{n-1\})$, where
$a:(0,\ldots,n-1)$, $b:(0,1)$, $c:{n-1 \choose 0}$, and $d: \one_Q$;
thus $\cU_n(a,b,c)=\cU_n(a,b,c,\emp)$.
Let $U_n(a,b,c,d)$ be the language of $\cU_n(a,b,c,d)$.

It was shown  by Cui, Gao, Kari, and Yu~\cite{CGKY_TCS12} that quaternary witnesses  meet the bound
$3\cdot 2^{m+n-2}-2^n+1$ for $(K_mL_n)^R$. They used witness $K_m$ with inputs (after relabelling) $a,b,c:\one_Q$, $d:(0,\ldots,m-1)$, and final state $m-1$, and witness $L_n$
with $a:(0,\ldots,n-1)$, $b:(n-2,n-1)$, $c:{n-1\choose n-2}$,  $d:\one_{Q_n}$ and final state $n-1$. Here $L_n$ is a dialect of $U_n(a,b,c,d)$.
We show that the  languages $U_m(a,b,c,d)$ and $U_n(d,c,b,a)$ also work.
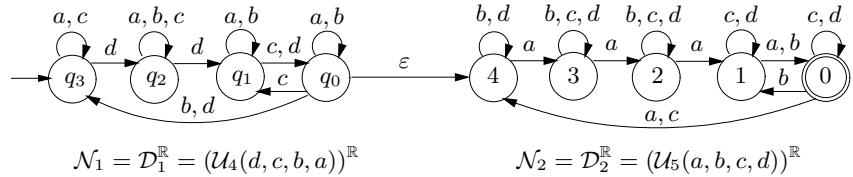
\begin{figure}[hbt]
\begin{center}
\setlength{\unitlength}{0.00039370in}
\begingroup\makeatletter\ifx\SetFigFont\undefined%
\gdef\SetFigFont#1#2#3#4#5{%
  \reset@font\fontsize{#1}{#2pt}%
  \fontfamily{#3}\fontseries{#4}\fontshape{#5}%
  \selectfont}%
\fi\endgroup%
{\renewcommand{\dashlinestretch}{30}
\begin{picture}(11084,2315)(0,-10)
\put(6680,109){\makebox(0,0)[lb]{\smash{{\SetFigFont{9}{10.8}{\rmdefault}{\mddefault}{\updefault}$\cN_2=\cD_2^\rev=(\cU_5(a,b,c,d))^\rev$}}}}
\put(1894.500,1705.929){\arc{394.717}{2.4948}{6.9299}}
\blacken\thicklines
\path(2054.033,1727.097)(2052.000,1587.000)(2125.998,1705.977)(2054.033,1727.097)
\thinlines
\put(4181.500,1713.929){\arc{394.717}{2.4948}{6.9299}}
\blacken\thicklines
\path(4341.033,1735.097)(4339.000,1595.000)(4412.998,1713.977)(4341.033,1735.097)
\thinlines
\put(806.500,1698.929){\arc{394.717}{2.4948}{6.9299}}
\blacken\thicklines
\path(966.033,1720.097)(964.000,1580.000)(1037.998,1698.977)(966.033,1720.097)
\thinlines
\put(7429.500,1721.929){\arc{394.717}{2.4948}{6.9299}}
\blacken\thicklines
\path(7589.033,1743.097)(7587.000,1603.000)(7660.998,1721.977)(7589.033,1743.097)
\thinlines
\put(6380.500,1706.929){\arc{394.717}{2.4948}{6.9299}}
\blacken\thicklines
\path(6540.033,1728.097)(6538.000,1588.000)(6611.998,1706.977)(6540.033,1728.097)
\thinlines
\put(8524.500,1713.929){\arc{394.717}{2.4948}{6.9299}}
\blacken\thicklines
\path(8684.033,1735.097)(8682.000,1595.000)(8755.998,1713.977)(8684.033,1735.097)
\thinlines
\put(9642.500,1728.929){\arc{394.717}{2.4948}{6.9299}}
\blacken\thicklines
\path(9802.033,1750.097)(9800.000,1610.000)(9873.998,1728.977)(9802.033,1750.097)
\thinlines
\put(10767.500,1713.929){\arc{394.717}{2.4948}{6.9299}}
\blacken\thicklines
\path(10927.033,1735.097)(10925.000,1595.000)(10998.998,1713.977)(10927.033,1735.097)
\thinlines
\put(817,1278){\ellipse{630}{630}}
\put(4164,1297){\ellipse{630}{630}}
\put(3035,1309){\ellipse{630}{630}}
\put(1900,1273){\ellipse{630}{630}}
\put(8522,1303){\ellipse{630}{630}}
\put(9643,1310){\ellipse{630}{630}}
\put(7429,1313){\ellipse{630}{630}}
\put(10758,1300){\ellipse{540}{540}}
\put(10761,1299){\ellipse{630}{630}}
\put(6372,1295){\ellipse{630}{630}}
\path(12,1287)(499,1287)
\blacken\thicklines
\path(379.000,1257.000)(499.000,1287.000)(379.000,1317.000)(379.000,1257.000)
\thinlines
\path(4504,1302)(6012,1302)
\blacken\thicklines
\path(5877.000,1264.500)(6012.000,1302.000)(5877.000,1339.500)(5877.000,1264.500)
\blacken\path(2608.000,1444.500)(2743.000,1482.000)(2608.000,1519.500)(2608.000,1444.500)
\thinlines
\path(2743,1482)(2158,1482)
\blacken\thicklines
\path(3763.000,1452.500)(3898.000,1490.000)(3763.000,1527.500)(3763.000,1452.500)
\thinlines
\path(3898,1490)(3313,1490)
\blacken\thicklines
\path(1520.000,1452.500)(1655.000,1490.000)(1520.000,1527.500)(1520.000,1452.500)
\thinlines
\path(1655,1490)(1070,1490)
\blacken\thicklines
\path(3417.000,1152.500)(3282.000,1115.000)(3417.000,1077.500)(3417.000,1152.500)
\thinlines
\path(3282,1115)(3912,1115)
\blacken\thicklines
\path(7048.000,1482.500)(7183.000,1520.000)(7048.000,1557.500)(7048.000,1482.500)
\thinlines
\path(7183,1520)(6598,1520)
\blacken\thicklines
\path(8143.000,1482.500)(8278.000,1520.000)(8143.000,1557.500)(8143.000,1482.500)
\thinlines
\path(8278,1520)(7693,1520)
\blacken\thicklines
\path(9231.000,1474.500)(9366.000,1512.000)(9231.000,1549.500)(9231.000,1474.500)
\thinlines
\path(9366,1512)(8781,1512)
\blacken\thicklines
\path(10055.000,1152.500)(9920.000,1115.000)(10055.000,1077.500)(10055.000,1152.500)
\thinlines
\path(9920,1115)(10505,1115)
\blacken\thicklines
\path(10356.000,1482.500)(10491.000,1520.000)(10356.000,1557.500)(10356.000,1482.500)
\thinlines
\path(10491,1520)(9906,1520)
\path(3919,1062)(3918,1062)(3916,1061)
	(3913,1059)(3907,1057)(3899,1053)
	(3888,1048)(3874,1042)(3857,1035)
	(3837,1026)(3814,1016)(3788,1005)
	(3759,993)(3727,980)(3693,966)
	(3657,952)(3619,937)(3579,922)
	(3537,907)(3494,891)(3449,876)
	(3403,861)(3356,846)(3308,831)
	(3258,817)(3207,803)(3154,790)
	(3099,777)(3042,765)(2983,754)
	(2922,744)(2858,734)(2792,725)
	(2723,718)(2652,712)(2578,707)
	(2503,704)(2427,702)(2347,702)
	(2269,705)(2193,709)(2120,715)
	(2050,722)(1983,730)(1920,740)
	(1859,750)(1801,762)(1746,774)
	(1693,787)(1642,800)(1593,814)
	(1545,829)(1499,844)(1455,859)
	(1412,874)(1371,890)(1331,906)
	(1294,921)(1258,936)(1224,950)
	(1192,964)(1164,977)(1137,989)
	(1114,1000)(1094,1009)(1077,1018)
	(1063,1024)(1052,1030)(1043,1034)(1031,1040)
\blacken\thicklines
\path(1168.518,1013.167)(1031.000,1040.000)(1134.977,946.085)(1168.518,1013.167)
\thinlines
\path(10625,1018)(10624,1018)(10623,1017)
	(10620,1016)(10615,1015)(10609,1012)
	(10600,1009)(10588,1005)(10574,1000)
	(10557,995)(10537,988)(10515,980)
	(10489,972)(10460,962)(10429,952)
	(10395,941)(10358,929)(10320,917)
	(10279,904)(10236,892)(10191,878)
	(10145,865)(10097,851)(10047,838)
	(9997,824)(9945,811)(9892,797)
	(9837,784)(9781,771)(9724,759)
	(9665,746)(9605,735)(9543,723)
	(9480,712)(9414,702)(9346,692)
	(9276,682)(9204,674)(9129,665)
	(9051,658)(8971,652)(8889,646)
	(8804,642)(8717,638)(8629,636)
	(8540,635)(8447,636)(8356,638)
	(8266,641)(8178,646)(8094,651)
	(8012,658)(7933,666)(7858,674)
	(7785,683)(7714,693)(7647,703)
	(7581,714)(7518,726)(7457,738)
	(7398,750)(7340,763)(7284,776)
	(7229,790)(7176,803)(7125,817)
	(7074,831)(7025,845)(6978,859)
	(6933,873)(6889,887)(6847,900)
	(6807,913)(6770,925)(6735,937)
	(6702,948)(6673,959)(6646,968)
	(6622,976)(6601,984)(6583,990)
	(6569,996)(6557,1000)(6547,1004)
	(6540,1006)(6530,1010)
\blacken\thicklines
\path(6669.272,994.680)(6530.000,1010.000)(6641.417,925.044)(6669.272,994.680)
\put(1790,1234){\makebox(0,0)[lb]{\smash{{\SetFigFont{9}{10.8}{\rmdefault}{\mddefault}{\updefault}$q_2$}}}}
\put(4084,1225){\makebox(0,0)[lb]{\smash{{\SetFigFont{9}{10.8}{\rmdefault}{\mddefault}{\updefault}$q_0$}}}}
\put(8367,725){\makebox(0,0)[lb]{\smash{{\SetFigFont{9}{10.8}{\familydefault}{\mddefault}{\updefault}$a,c$}}}}
\put(6289,1223){\makebox(0,0)[lb]{\smash{{\SetFigFont{9}{10.8}{\rmdefault}{\mddefault}{\updefault}$4$}}}}
\put(7347,1230){\makebox(0,0)[lb]{\smash{{\SetFigFont{9}{10.8}{\rmdefault}{\mddefault}{\updefault}$3$}}}}
\put(9559,1231){\makebox(0,0)[lb]{\smash{{\SetFigFont{9}{10.8}{\rmdefault}{\mddefault}{\updefault}$1$}}}}
\put(10676,1231){\makebox(0,0)[lb]{\smash{{\SetFigFont{9}{10.8}{\rmdefault}{\mddefault}{\updefault}$0$}}}}
\put(7850,1657){\makebox(0,0)[lb]{\smash{{\SetFigFont{9}{10.8}{\familydefault}{\mddefault}{\updefault}$a$}}}}
\put(8973,1656){\makebox(0,0)[lb]{\smash{{\SetFigFont{9}{10.8}{\familydefault}{\mddefault}{\updefault}$a$}}}}
\put(5127,1423){\makebox(0,0)[lb]{\smash{{\SetFigFont{9}{10.8}{\familydefault}{\mddefault}{\updefault}$\eps$}}}}
\put(716,1220){\makebox(0,0)[lb]{\smash{{\SetFigFont{9}{10.8}{\rmdefault}{\mddefault}{\updefault}$q_3$}}}}
\put(1234,1602){\makebox(0,0)[lb]{\smash{{\SetFigFont{9}{10.8}{\familydefault}{\mddefault}{\updefault}$d$}}}}
\put(2381,1610){\makebox(0,0)[lb]{\smash{{\SetFigFont{9}{10.8}{\familydefault}{\mddefault}{\updefault}$d$}}}}
\put(2922,1248){\makebox(0,0)[lb]{\smash{{\SetFigFont{9}{10.8}{\rmdefault}{\mddefault}{\updefault}$q_1$}}}}
\put(3380,1616){\makebox(0,0)[lb]{\smash{{\SetFigFont{9}{10.8}{\familydefault}{\mddefault}{\updefault}$c,d$}}}}
\put(3521,1197){\makebox(0,0)[lb]{\smash{{\SetFigFont{9}{10.8}{\familydefault}{\mddefault}{\updefault}$c$}}}}
\put(2247,830){\makebox(0,0)[lb]{\smash{{\SetFigFont{9}{10.8}{\familydefault}{\mddefault}{\updefault}$b,d$}}}}
\put(560,2014){\makebox(0,0)[lb]{\smash{{\SetFigFont{9}{10.8}{\familydefault}{\mddefault}{\updefault}$a,c$}}}}
\put(1555,2013){\makebox(0,0)[lb]{\smash{{\SetFigFont{9}{10.8}{\familydefault}{\mddefault}{\updefault}$a,b,c$}}}}
\put(3951,2006){\makebox(0,0)[lb]{\smash{{\SetFigFont{9}{10.8}{\familydefault}{\mddefault}{\updefault}$a,b$}}}}
\put(8456,1223){\makebox(0,0)[lb]{\smash{{\SetFigFont{9}{10.8}{\rmdefault}{\mddefault}{\updefault}$2$}}}}
\put(9965,1653){\makebox(0,0)[lb]{\smash{{\SetFigFont{9}{10.8}{\familydefault}{\mddefault}{\updefault}$a,b$}}}}
\put(6138,2040){\makebox(0,0)[lb]{\smash{{\SetFigFont{9}{10.8}{\familydefault}{\mddefault}{\updefault}$b,d$}}}}
\put(2818,2021){\makebox(0,0)[lb]{\smash{{\SetFigFont{9}{10.8}{\familydefault}{\mddefault}{\updefault}$a,b$}}}}
\put(10513,2030){\makebox(0,0)[lb]{\smash{{\SetFigFont{9}{10.8}{\familydefault}{\mddefault}{\updefault}$c,d$}}}}
\put(9410,2037){\makebox(0,0)[lb]{\smash{{\SetFigFont{9}{10.8}{\familydefault}{\mddefault}{\updefault}$c,d$}}}}
\put(8119,2037){\makebox(0,0)[lb]{\smash{{\SetFigFont{9}{10.8}{\familydefault}{\mddefault}{\updefault}$b,c,d$}}}}
\put(7046,2045){\makebox(0,0)[lb]{\smash{{\SetFigFont{9}{10.8}{\familydefault}{\mddefault}{\updefault}$b,c,d$}}}}
\put(6761,1648){\makebox(0,0)[lb]{\smash{{\SetFigFont{9}{10.8}{\familydefault}{\mddefault}{\updefault}$a$}}}}
\put(10129,1193){\makebox(0,0)[lb]{\smash{{\SetFigFont{9}{10.8}{\familydefault}{\mddefault}{\updefault}$b$}}}}
\put(845,110){\makebox(0,0)[lb]{\smash{{\SetFigFont{9}{10.8}{\rmdefault}{\mddefault}{\updefault}$\cN_1=\cD_1^\rev=(\cU_4(d,c,b,a))^\rev$}}}}
\thinlines
\put(3041.500,1735.929){\arc{394.717}{2.4948}{6.9299}}
\blacken\thicklines
\path(3201.033,1757.097)(3199.000,1617.000)(3272.998,1735.977)(3201.033,1757.097)
\end{picture}
}
\end{center}
\caption{NFA for $(U_4(d,c,b,a))^R\;(U_5(a,b,c,d))^R$.} 
\label{fig:KrLr}
\end{figure}

\begin{theorem}[$L_n^RK_m^R$]
The complexity of $(U_n(d,c,b,a))^R(U_m(a,b,c,d))^R$ is $3\cdot 2^{m+n-2} -2^n+1$, for $m,n \ge 3$.
\end{theorem}
\begin{proof}
Let $\cD_1 = (Q_1, \Sigma, \delta_1, q_0, \{q_{n-1}\})$, where  $Q_1=\{q_0,\ldots,q_{n-1}\}$, and $\cD_2 = (Q_2, \Sigma, \delta_2, 0, \{m-1\})$, where $Q_2=\{0,\ldots,{m-1}\}$, be the minimal DFA's of $U_n(d,c,b,a)$  and $U_m(a,b,c,d)$.
Let $\cN_1=\cD_1^\rev$, $\cN_2=\cD_2^\rev$, and let $\cN$ be the NFA for the product of $\cN_1$ and $\cN_2$, as illustrated in Fig.~\ref{fig:KrLr}.
We use the subset construction to get a DFA $\cP$ for this product.
Any reachable state of $\cP$ must either not contain $q_0$ or contain both $q_0$ and $n-1$.
We will show that all $2^{n+m-1}$ and $2^{n+m-2}$ states of these two forms are reachable.

The initial state of $\cP$ is $\{q_{n-1}\}$. 
It is known from \cite{BrTa12,SWY04} that all $2^n$ subsets of $Q_1$ are reachable in $\cN_1$
by words in $\{b,c,d\}^*$.
Of these inputs, $b$ and $d$ map state $\{m-1\}$ of $\cN_2$ to itself, and $c$ maps $\{m-1\}$  to $\emp$.
Suppose state $S \subseteq Q_1$ is reached by applying the word $w \in \{b,c,d\}^*$ to $\cN_1$.
If $q_0 \in S$, then the state $S \cup \{q_{m-1}\}$ is reachable in $\cP$ by $w$.
If $q_0 \notin S$,  since $c^2$ is the identity transformation on $\cN_1$, 
state $S$ of $\cP$ is reachable by $wc^2$.

We have the chain $\{q_{n-1}\} \der{d^n} \{q_{n-1}, m-1\} \der{b} \{m-1\}$. 
In a way similar to that in  $\cN_1$, all $2^m$ subsets of $Q_2$ are reachable in $\cN_2$ by words in $\{a,b,c\}^*$.
Applying the same words to $\{m-1\}$ in $\cP$ yields all  subsets of~$Q_2$.

Now suppose that all states of the form $S \cup T$, $S \subseteq Q_1 \bs \{q_0\}$, $T \subseteq Q_2$, $|T| = k \ge 0$ are reachable.
We will show that all states of the form $S \cup T$, $|T| = k+1$ are reachable.
Let $T = \{t_1, t_2, \dots, t_{k+1}\}$, $t_1 < \cdots < t_{k+1}$. 
Let $S \subseteq Q_1 \bs \{q_0\}$. 
We have already established that if $S = \emp$, $S \cup T$ is reachable.
Otherwise, define $j = m-1 - t_{k+1}$, and $T' = \{t_1 + j, t_2 + j, \dots, t_k + j\}$.
Then $S \cup T' \der{d^n} S \cup T' \cup \{m-1\} \der{a^j} S \cup T$.

Now suppose $q_0 \in S$. 
If $S \neq Q_1$, there exists an $S'$ with $q_0 \notin S'$ such that $S' \cup T \der{a^j} S \cup T$ for some $j$.
Note that $S' \cup T$ is reachable by the previous case.
If $S = Q_1$, then define $S' = Q_1 \bs \{q_0\}$.
Once again, $S' \cup T$ is reachable, and we have $S' \cup T \der{db^2} S \cup T$.

Therefore all $2^{n+m-1} + 2^{n+m-2} = 3 \cdot 2^{n+m-2}$ states not containing $q_0$ or containing both $q_0$ and $m-1$ are reachable.

For distinguishability, first note that all $2^n$ states of the form $S \cup Q_2$ are final and indistinguishable.
Consider a pair of states $S_1 \cup T_1$, $S_2 \cup T_2$, with $S_1, S_2 \subseteq Q_1$ and $T_1, T_2 \subsetneq Q_2$. 
If $T_1 \neq T_2$, let $k \in T_1 \oplus T_2$; then $a^k$ distinguishes the states.
Otherwise, $T_1 = T_2 = T$, and $S_1 \neq S_2$. 
Since $T \neq Q_2$, there exists a $k \notin T$.
Also, there exists $q_l \in S_1 \oplus S_2$.
Applying $d^la^{k+1}$ results in states $S_1' \cup T_1'$, $S_2' \cup T_2'$ such that $m-1 \in T_1' \oplus T_2'$.
Therefore all remaining states are distinguishable by using the previous argument.
\qed
\end{proof}

\subsection{Reverse of Star}
Note that $(L^*)^R=(L^R)^*$.
The star of the reverse was studied by Gao, K.~Salomaa, and Yu~\cite{GSY_FI08}, who showed that the complexity of this operation is $2^n$.
The witness they used is a dialect of $U_n(a,b,c)$. 
After relabelling of states and permuting the inputs, it has the following transformations:
$a:(0,\ldots,n-1)$, $b:(0,n-1)$ and $c:{0\choose n-1}$, and the final state is 0.
The witness $\cU_{\{0\},n}(a,b,c)$, which is $\cU_n(a,b,c)$ with final state set changed to $\{0\}$ also works, as does every dialect of $\cU_n(a,b,c)$ with final state set $\{0\}$.
\begin{theorem}[$(L^*)^R$]
\label{thm:LRstar}
For $n \ge 3$,  the  complexity of $((U_{\{0\},n}(a,b,c))^*)^R$ is $2^n$.
\end{theorem}
\begin{proof}
The proof is the same as that in~\cite{GSY_FI08}. Since $L_n$ has only one final state which is also the initial state, we have $L_n^*=L_n$. Hence $(L_n^*)^R=L_n^R$, and $L_n^R$ has state complexity $2^n$.
\qed
\end{proof}

\section{Conclusions}
\label{sec:conc}

We have proved that the universal witnesses $U_n(a,b,c)$ and  $U_n(a,b,c,d)$, along with their permutational equivalents $U_n(b,a,c)$ and $U_n(d,c,b,a)$, and dialects
$U_{\{0,2\},m}(a,b,c)$, $U_{\{1,3\},n}(a,b,c)$, $U_{\{0\},n}(a,b,c)$, $V_m(a,b,c,d)$ and
$V_n(d,c,b,a)$ suffice to act as witnesses for all the state complexity bounds involving binary boolean operations, product, star and reversal.
We have shown that it is efficient to consider all four boolean operations together.
Lastly, the use of universal witnesses and their dialects simplified many proofs, and allowed us to utilize the similarities in the witnesses.
\medskip

\noin 
{\bf Acknowledgment} We thank Baiyu Li for careful proofreading.

\end{document}